\documentclass[10pt, aps,twocolumn,prc,floatfix,preprintnumbers,superscriptaddress,notitlepage,nofootinbib]{revtex4-1}
\usepackage{dcolumn}
\usepackage{bm}
\usepackage{color}
\usepackage{amssymb}
\usepackage{amsmath}
\usepackage{graphicx}
\usepackage{amsfonts}
\usepackage{slashed}
\usepackage{pstricks}
\usepackage{float}
\usepackage[colorlinks = true,
            linkcolor = blue,
            urlcolor  = blue,
            citecolor = blue,
            anchorcolor = blue]{hyperref}
\usepackage[capitalize]{cleveref}
\usepackage{array}
\allowdisplaybreaks
\usepackage{dcolumn}
\usepackage{epsf}
\usepackage{slashed}
\usepackage[compat=1.1.0]{tikz-feynman}
\usepackage{tikz}
\usetikzlibrary{shapes.geometric}
\usetikzlibrary{calc}
\usepackage{orcidlink}
\usepackage{xspace}
\usepackage{MnSymbol}
\usepackage{multirow}
\usepackage{soul}

\newcommand{\bra}[1]{\ensuremath{\langle #1 |}}   
\newcommand{\ket}[1]{\ensuremath{| #1 \rangle}}   

\DeclareMathOperator*{\SumInt}{%
\mathchoice%
  {\ooalign{$\displaystyle\sum$\cr\hidewidth$\displaystyle\int$\hidewidth\cr}}
  {\ooalign{\raisebox{.14\height}{\scalebox{.7}{$\textstyle\sum$}}\cr\hidewidth$\textstyle\int$\hidewidth\cr}}
  {\ooalign{\raisebox{.2\height}{\scalebox{.6}{$\scriptstyle\sum$}}\cr$\scriptstyle\int$\cr}}
  {\ooalign{\raisebox{.2\height}{\scalebox{.6}{$\scriptstyle\sum$}}\cr$\scriptstyle\int$\cr}}
}

\definecolor{gray}{rgb}{0.6,0.6,0.6}
\definecolor{darkgreen}{rgb}{0.0, 0.545098, 0.0}
\definecolor{darkblue}{rgb}{0.0, 0.0, 0.545098}
\definecolor{BrickRed}{rgb}{0.8, 0.25, 0.33}
\definecolor{gray}{rgb}{0.6,0.6,0.6}
\definecolor{darkgreen}{rgb}{0.0, 0.545098, 0.0}
\definecolor{mypink1}{rgb}{0.858, 0.188, 0.478}

\def\Fermilab{Theory Division, Fermilab, P.O. Box 500, Batavia, IL 60510, USA}

\newcommand{\achilles}{\textsc{Achilles}\xspace}

\begin{document}
\preprint{FERMILAB-PUB-25-0473-T}
\preprint{MSUHEP-25-018}
\title{Single pion-production and pion propagation in \achilles}

\author{Joshua Isaacson\,\orcidlink{0000-0001-6164-1707}}
\email{isaacs21@msu.edu}
\affiliation{Department of Physics and Astronomy, Michigan State University, East Lansing, MI 48824, USA}
\affiliation{\Fermilab}

\author{William Jay\,\orcidlink{0000-0002-5918-222X}}
\email{william.jay@colostate.edu}
\affiliation{Department of Physics, Colorado State University, Fort Collins, CO 80523, USA}

\author{Alessandro Lovato\,\orcidlink{0000-0002-2194-4954}}
\email{lovato@anl.gov}
\affiliation{Physics Division, Argonne National Laboratory, Argonne, Illinois 60439, USA}
\affiliation{Computational Science Division, Argonne National Laboratory, Argonne, Illinois 60439, USA}
\affiliation{INFN-TIFPA Trento Institute for Fundamental Physics and Applications, Trento, Italy}

\author{Pedro~Machado\,\orcidlink{0000-0002-9118-7354}}
\email{pmachado@fnal.gov}
\affiliation{\Fermilab}

\author{Alexis Nikolakopoulos\,\orcidlink{0000-0001-6963-8115}}
\email{anikolak@uw.edu}
\affiliation{University of Washington, Seattle, WA 98195, USA}

\author{Noemi Rocco\,\orcidlink{0000-0002-7150-7322}}
\email{nrocco@fnal.gov}
\affiliation{\Fermilab}

\author{Noah Steinberg\,\orcidlink{0000-0003-0427-4888}}
\email{nsteinberg@anl.gov}
\affiliation{Physics Division, Argonne National Laboratory, Argonne, Illinois 60439, USA}
\affiliation{\Fermilab}

\begin{abstract}
We extend the applicability of \achilles (A CHIcagoLand Lepton Event Simulator) by incorporating the single-pion production mechanism in a fully exclusive fashion. 
The electroweak interaction vertex is modeled by combining the state-of-the-art Dynamical Coupled-Channels approach with realistic hole spectral functions, which account for correlations in both the initial target state and the residual spectator system. 
Final-state interactions are treated using a semi-classical intranuclear cascade that leverages nuclear configurations to determine the correlated spatial distribution of protons and neutrons. 
The meson-baryon scattering amplitudes used in the cascade are computed within the Dynamical Coupled-Channels framework, consistent with the electroweak vertex. 
To model pion absorption, we employ the optical potential approach of Oset and Salcedo. 
As an alternative approach, we explicitly model the production and propagation of resonances which mediate pion-nucleon scattering and pion absorption.
We validate our approach against pion-nucleon and pion-nucleus scattering data, and present comparisons with electron- and neutrino–nucleus measurements from e4$\nu$, T2K, MINER$\nu$A, and MicroBooNE. 
\end{abstract}

\maketitle


\section{Introduction}\label{sec:intro}
Accurately modeling neutrino–nucleus scattering in the resonance-production region is critical for the success of accelerator-based neutrino oscillation programs. This region is particularly important for Deep Underground Neutrino Experiment (DUNE)~\cite{DUNE:2020jqi}, as the incoming neutrino flux spans energies from 1 to 10 GeV. A precise understanding of resonance production is also essential for experiments with lower-energy fluxes, such as the planned Hyper-Kamiokande (HK)~\cite{Hyper-Kamiokande:2018ofw}, since pions produced in neutrino interactions can be reabsorbed within the nucleus, causing such events to be misidentified as quasi-elastic. This effect is especially relevant when comparing predictions from nuclear \textit{ab initio} methods to flux-folded cross sections, as these misidentified events must be excluded from the measured cross section~\cite{Lovato:2020kba,Acharya:2024xah}.

Capitalizing on liquid-argon detectors---such as those used in the Short Baseline Neutrino (SBN) program~\cite{AlvarezGarrote:2024szs} and DUNE---requires modeling neutrino–nucleus scattering in an exclusive manner. A complete quantum-mechanical treatment of the complex nuclear dynamics involved in this process will remain unfeasible in the foreseeable future, despite ongoing progress using quantum computers to simulate simplified scenarios~\cite{Roggero:2019myu,Weiss:2024mie}. Neutrino event generators make the problem tractable by separating the electroweak interaction vertex—treated at the quantum-mechanical level—from the subsequent intranuclear cascade (INC), which models the semi-classical propagation of hadrons through the nuclear medium~\cite{Andreopoulos:2009rq,Golan:2012rfa,Mosel:2016cwa,Hayato:2021heg}.

This work focuses on single-pion production on nuclei, a multi-scale problem that involves both hadronic and nuclear excitations in a region where quantum chromodynamics is strongly coupled. In addition, the momenta at play are much larger than the pion mass, which places it outside the radius of convergence of chiral effective field theory in this regime. 
Thus, the model developed by Hern\'andez, Nieves, and Valverde (HNV)\cite{Hernandez:2007qq}, which evaluates the non-resonant background amplitudes from a leading-order chiral Lagrangian, is limited at intermediate energies. To extend its applicability, particularly in the $\Delta$ region, the authors of Ref.~\cite{Alvarez-Ruso:2015eva} have partially unitarized the model by imposing Watson’s
theorem to the dominant vector and axial multipoles.

Two of the most widely used event generators, NEUT and GENIE, rely on the Rein–Sehgal model~\cite{Rein:1980wg} and its Berger–Sehgal extension~\cite{Berger:2007rq}. These frameworks simulate pion production through the incoherent excitation of multiple nucleon resonances—primarily the $\Delta(1232)$, but also higher-mass $N^*$ and $\Delta$ states—followed by their decay into a nucleon and a pion. The Berger–Sehgal model improves upon the original formalism by accounting for the charged lepton mass and refining the pion angular distributions. Vector transition form factors are constrained by electroproduction data, while axial contributions are typically modeled with a dipole form and tuned to historical neutrino–deuteron data~\cite{Wilkinson:2014yfa}. NuWro has long included only the $\Delta(1232)$ explicitly and treated higher invariant mass production via a smooth interpolation to a Deep-Inelastic-Scattering (DIS)-based hadronization model, invoking quark-hadron duality~\cite{Golan:2012wx}. More recently, NuWro has adopted the Ghent hybrid model~\cite{Yan:2024kkg}. In the low-energy region, the resonant and non-resonant contributions are those of the HNV model. At higher energies, the description of the non-resonant background is instead based on Regge phenomenology~\cite{Gonzalez-Jimenez:2016qqq,Guidal:1997hy}.

In this work, we implement in \achilles the Dynamical Coupled-Channels (DCC) model~\cite{Kamano:2013iva,Nakamura:2015rta,Kamano:2016bgm,Doring:2025sgb} to compute the single-pion production amplitudes. 
The DCC approach provides a unified, unitary description of meson–baryon scattering and electroweak pion production by solving coupled integral equations for multiple reaction channels.
This framework enables a consistent treatment of both resonant and non-resonant contributions, including their interference, across a broad range of energies and final states. 
We note that the recent implementation of the DCC model in NEUT~\cite{Yamauchi:2024lby} is fully inclusive. Therefore, it computes the hadronic response via the structure functions, and is therefore limited to providing only lepton kinematics, without access to the hadronic final state. 
In contrast, our implementation retains the full dynamical information of the DCC amplitudes, allowing for the exclusive modeling of the hadronic final state, including pion and nucleon kinematics.

To account for correlations in the initial target state and the residual spectator system, we employ realistic hole spectral functions of $^{12}$C~\cite{Rocco:2019gfb} and $^{40}$Ar~\cite{Nikolakopoulos:2024mjj} nuclei. 
After the primary interaction, all produced hadrons are propagated through the nuclear medium using an INC that accounts for scattering, absorption, and charge exchange. 
To gauge the sensitivity to a range of in-medium effects, we implement and test two different cascade modes depending on whether the $\Delta$ and higher nucleon resonances are propagating degrees of freedom. 

In the first approach, the relevant meson–baryon cross sections are computed from the standard decomposition of the scattering matrix into partial-wave amplitudes. 
Crucially, to ensure consistency with the electroweak interaction vertex, these amplitudes are computed within the Dynamical Coupled-Channels (DCC) framework. 
In addition to meson–baryon scattering, where a meson remains in the final state, absorption processes play a significant role in the INC. Here, we implement an approach commonly used in neutrino event generators, based on the optical potential model of Oset and Salcedo~\cite{Salcedo:1987md}, in which mesons are absorbed via the imaginary part of their self-energy in nuclear matter. Crucially in this approach, nucleon resonances are not propagating degrees of freedom.

In the second approach, we follow the approach taken by GiBUU~\cite{Buss:2011mx} and INCL~\cite{Cugnon:1996kh} where we treat the $\Delta$ resonance as a dynamical degree of freedom that propagates throughout the nuclear medium. This approach provides a more microscopic description of three-body processes involved in pion absorption (e.g., $\pi N \leftrightarrow \Delta, \Delta N\leftrightarrow NN$), and accurately accounts for the relatively long lifetime of the $\Delta$ resonance, compared to typical intranuclear propagation timescales.
This provides a mechanism to capture re-scattering effects before the decay of the $\Delta$ and for medium modifications due to nuclear correlations.

This paper is organized as follows. In Sect.~\ref{sec:theory}, we describe both the electroweak interaction vertex—including how the DCC model is incorporated into the extended factorization scheme—and the INC, using the two different cascade modes. Sect.~\ref{sec:validation} presents a validation of our approach, beginning with inclusive electron-nucleus scattering, followed by comparisons to pion–nucleus scattering observables. Comparisons with exclusive electron- and neutrino–nucleus scattering data, including results from the e4$\nu$, T2K, MINER$\nu$A, and MicroBooNE experiments, are presented in Section~\ref{sec:compare_e_nu}. Finally, in Section~\ref{sec:conclusions}, we summarize our conclusions and outline future directions for \achilles development.

\section{Theoretical Treatment}
\label{sec:theory}
\achilles relies upon a factorization of lepton-nucleus scattering into an incoherent product of the initial electroweak interaction, and the subsequent intranuclear cascade which transports particles out of the nucleus~\cite{Isaacson:2022cwh}. This amounts to approximating the squared matrix element as
\begin{align}\label{eq:genfact}
    |\mathcal{M}(\{k\} \rightarrow\{p^{\prime}\})|^{2} \simeq &\SumInt_{p}|\mathcal{V}(\{k\} \rightarrow \{p\})|^{2} \\ \nonumber
    & \quad\times |\mathcal{P}(\{p\} \rightarrow \{p^{\prime}\})|^{2}\,.
\end{align}
In \cref{eq:genfact}, $\mathcal{V}(\{k\} \rightarrow \{p\})$ is the amplitude for the initial hard interaction process, which in \achilles is provided by the spectral-function formalism. The intranuclear cascade (INC) in \achilles provides $\mathcal{P}(\{p\} \rightarrow \{p^{\prime}\})$, the probability for particles $\{p^{\prime}\}$ to leave the nucleus when starting with particles $\{p\}$ inside the nucleus. The INC takes as input nuclear configurations and uses a semiclassical, impact-parameter-based algorithm to model the propagation of particles in the nuclear medium~\cite{Isaacson:2020wlx}. Below we describe the ingredients necessary for both the hard interaction and the INC.

\subsection{ANL-Osaka dynamical coupled-channel approach \label{sec:DCC}}

The ANL-Osaka dynamical model describes $\pi N$ and $\gamma N$ reactions in terms of hadron degrees of freedom by solving a coupled-channel scattering equation, which preserves unitarity of scattering amplitudes.
More explicitly, the approach employs an energy-independent Hamiltonian of the form
\begin{align}
H &= H_0 + \Gamma + v. \label{eq:HamiltonianDCC}
\end{align}
The first term $H_0$ corresponds to the free Hamiltonian.
The second term $\Gamma$ describes production and decay of ``bare" excited-hadron states such as the $N^\ast$ and $\Delta$ from a meson-baryon channel (e.g., $\pi N$ or $\eta N$).
The third term $v$ describes non-resonant interactions as a sum of energy-independent meson-exchange potentials which are derived by applying a unitarity transformation to a  phenomenological Lagrangian~\cite{Sato:1996gk,1997PThPh..98..927K}.  

The Hamiltonian in \cref{eq:HamiltonianDCC} 
contains stable asymptotic particles as well as unstable particles.
What is desired is an effective description of the scattering process for the former that consistently includes the influence of the latter.
Standard Feshbach operator projection techniques~\cite{Feshbach:1992zp} can be applied to produce an energy-dependent non-local Hamiltonian in terms of the stable particles only.
The problem at hand is phrased most conveniently in terms of two-body scattering matrix $T$.
In Ref.~\cite{Matsuyama:2006rp}, operator projection techniques were applied to \cref{eq:HamiltonianDCC} to yield a unitary set of coupled-channel equations, allowing the $T$-matrix to be written as
\begin{align}
\langle \beta |T^\pm(s) |\alpha \rangle = \langle \beta | t^\pm_{\rm bg}(s) |\alpha \rangle
+ \langle \beta |t^\pm_{\rm res}(s) |\alpha \rangle
\label{eq:dcc_t-matrix}
\end{align}
where $s$ is the square of the scattering center-of-mass energy and $\alpha, \beta$ denote meson-baryon channels.
Partial contributions to the total $T$ matrix are denoted with lower-case $t$. The  superscript $\pm$ indicates
if the outgoing/incoming boundary  condition is applied; in the following equations, we omit this superscript for brevity.
We have separated the resonant contribution to the $T^\pm$ matrix, $t_{\rm res}^\pm$, from the non-resonant background contribution, $t_{\rm bg}^\pm$.

The non-resonant meson-baryon $t$-matrix is obtained by solving the coupled-channel equation as 
\begin{align}
\langle \beta |t_{\rm bg}(s) |\alpha \rangle= \langle \beta | V(s) + V(s) G_{\rm MB}(s) t_{\rm bg}(s) |\alpha \rangle
\end{align}
where $G_{\rm MB}(s)$ is the Green's function of a meson-baryon state.
The effective two-body interaction $V(s)$ is the sum of the non-resonant interaction $v$ and the contribution of the so-called ``Z-diagram" mechanisms which include $\pi \pi N$ intermediate states, for more details see Fig. 8 in Ref.~\cite{Matsuyama:2006rp}.

The resonant $t$-matrix is given by the sum of the $N^\ast$ or $\Delta$ resonances as
\begin{align}
\langle \beta | t_{\rm res}(s) |\alpha \rangle = \sum_{m,n} \langle \chi_\beta |\Gamma |N^\ast_m\rangle [D]_{m,n} \langle N_m^\ast |\Gamma |\chi_\alpha \rangle 
\end{align}
where $[D]_{m,n}$ is the resonance propagator and $\langle N_m^\ast |\Gamma |\chi_\alpha \rangle $ describes the production vertex of the resonance state from a distorted meson-baryon state
\begin{align}
|\chi_\alpha \rangle = \left[1+ G_{\rm MB}(s) t_{\rm bg}(s) \right]|\alpha \rangle,
\end{align}
where the distortion on the right-hand side arises from the non-resonant interaction.
All free parameters of the meson-exchange interaction $v$ are determined by fitting the data for $\pi N \rightarrow \pi N$ from the SAID database~\cite{cns_dac}. 
Ref.~\cite{Kamano:2013iva} gives a detailed discussion of how the parameters and masses entering in the $N^\ast$ transition to meson-baryon states are determined using different datasets. 

Within the DCC model, the electroweak pion-production matrix element which provides the hard matrix element of \cref{eq:genfact}, is a sum of non-resonant and resonant contributions
\begin{align}\label{eq:piNcurrent}
\langle \alpha|j^\mu|N\rangle = &\,\langle \chi_\alpha | j^\mu_{\rm bg}|N\rangle \\ & \quad + \sum_{m,n} \langle \chi_\alpha |\Gamma |N_m^\ast \rangle [D]_{m,n} \langle N_m^\ast | j^\mu | N\rangle,\nonumber
\end{align}
where $\langle\alpha|=\langle\pi N|$ is a pion-nucleon final state, and the first and second terms on the right hand side of \cref{eq:piNcurrent} are the electroweak counterparts of $v$ and $\Gamma$ in \cref{eq:HamiltonianDCC}, respectively.
The current in the resonant amplitude $\langle N_m^\ast |j^\mu |N \rangle$ can be further decomposed as
\begin{align}
j^\mu = j^\mu_{\rm res}+\Gamma\left[1+t_{\rm bg}(s)\right]G_{\rm MB}(s)j^\mu_{\rm bg}.
\end{align}
The first term in the resonant amplitude comes from the direct coupling of the current to the resonance in the DCC model, while the second term arises because a re-scattering process can also excite the resonance.

The dynamical model was initially developed to study the electromagnetic structure of the $\Delta$ resonance. In this energy regime, only the $\pi N$ channel is open. The vector transition form factors at finite momentum transfer $Q^2$ are extracted from analyses of pion electroproduction in Ref.~\cite{DCC1,DCC2}.
To extend the model to neutrino-induced reactions, Ref.~\cite{DCC2} introduced
an axial-vector current component, guided by the partially conserved axial current (PCAC) hypothesis,\footnote{The PCAC hypothesis refers to usage of the axial Ward identity together with the assumption of pion pole dominance at low $Q^2$ to write the matrix element for $\pi^+ n \to R^+$, with $R^+$ a resonance, as $\bra{R^+} \partial_\mu A^\mu(0)\ket{n} = -i m_\pi^2 f_\pi \frac{1}{q^2-m_\pi^2} T(\pi^+ n \to R^+)$ where $T$ is the on-shell pion-nucleon matrix element.}
while all parameters associated with the strong interaction and the vector current are kept unchanged.

The ANL-Osaka DCC amplitudes given in~\cref{eq:piNcurrent} are expanded in partial waves and provide the basis of our calculations for single pion production in electron- and neutrino-nucleus scattering. Amplitudes for electromagnetic (EM), charged current (CC), and neutral current (NC) processes are provided in tabularized form in the $\pi N$ center-of-mass frame and boosted to the lab frame following the procedure given in Ref.~\cite{Rocco:2019gfb}. A more thorough discussion of their implementation is given in the next section.
The DCC model has been successfully extended to account for the production of $\pi\pi N$ final states~\cite{DCC1}.
However, discussion of these developments exceeds the scope of the present work.

\subsection{Electroweak Interaction Vertex}\label{sec:EIV}

As mentioned above and emphasized in Ref.~\cite{Isaacson:2022cwh}, the theoretical treatment of the hard scattering begins with an explicit factorization ansatz for the hadronic final state $\ket{\Psi_f}$ immediately following the initial electroweak interaction. In the quasi-elastic region, the nuclear current operator reduces to a sum of one-body terms and the nuclear final state factorizes as $|\Psi_f^A \rangle = |p\rangle \otimes |\Psi_f^{A-1}\rangle$. Here, $|p\rangle$ denotes the final-state nucleon with momentum ${\bf p}$ and energy $e({\bm p})=\sqrt{{\bm p}^2+m_N^2} $, while $|\psi_f^{A-1}\rangle$ describes the \hbox{$(A-1)$-body} spectator system. Supposing a momentum transfer of $q^\mu=(\omega, \bm{q})$, the energy and momentum of the recoiling system are fixed by energy and momentum conservation
\begin{align}
E_f^{A-1}=\omega +E_0-e({\bm p})\, ,\quad {\bm P}^{A-1}_f={\bm q}-{\bm p}\, ,
\end{align}
where $E_{0}$ is the ground state energy of the initial state nucleus.
The incoherent contribution to the hadron tensor with one nucleon (1N) in the final state is given by 
\begin{align}
W^{\mu\nu}_{\rm 1N}(p,q,s_p,t_p)=
& \sum_{s_k,t_k} \int \frac{d^3k}{(2\pi)^3} \frac{1}{2e({\bm k})}dE S_{t_k}({\bm k},E)\nonumber\\
&\times  \langle k | {j^\mu}^\dagger |p \rangle \langle p |  j^\nu | k \rangle\, \delta^{(3)}(\bm{p} - \bm{k} - \bm{q}) \nonumber\\
& \times \delta(\omega-E+m_N-e(\bm{k}+\bm{q})) \,,
\label{had:tens}
\end{align}
where the integration runs over the removal energy and initial momentum of the initial-state nucleon and where $s_k$, $t_k$ ($s_p$, $t_p$) label the spin
and isospin of the initial (final) nucleons, $m_N$ is the rest mass of the nucleon. We note that, consistent with Ref.~\cite{Isaacson:2022cwh}, the phase space integrals over the momenta and the spin-isospin summations of the final-state particles are performed outside the hadron tensor, (see App.~\ref{app:phase_space} for details on the phase space integrals). \achilles performs the summation over these quantum numbers, convolutes the result with the one-body spectral function, and enforces energy and momentum conservation. The hole spectral function $S_{t_k}(\bm{k},E)$ provides the probability distribution of removing an initial-state nucleon with momentum ${\bm k}$ from the target nucleus, leaving the residual $(A-1)$-nucleon system with an excitation energy $E$; it is normalized to the number of protons $Z$ and the number of neutrons $A-Z$, depending on $t_k$.

The factorization scheme readily generalizes to include a pion in the final state~\cite{Rocco:2019gfb}:
\begin{align}
|\Psi_f^A\rangle \rightarrow |p_\pi p\rangle \otimes |\Psi_f^{A-1}\rangle \, ,
\end{align}
where a nucleon and a pion with four-momenta $p$ and $p_\pi$ are produced in the hard interaction, respectively. The incoherent contribution to the hadron response tensor with one nucleon and one pion (1N1$\pi$) in the continuum can be written as~\cite{Rocco:2019gfb}
\begin{align}
& W^{\mu\nu}_{\rm 1N 1\pi}(p,q,s_p,t_p,p_\pi) = \sum_{s_k, t_k} \int \frac{d^3k}{(2\pi)^3} dE \frac{1}{4e({\bf k})e({\bf p_\pi})}\nonumber\\
&\times P_{t_k}({\mathbf k},E) \langle k | {j^\mu}^\dagger |p_\pi p\rangle \langle p_\pi p |  j^\nu | k \rangle\, \delta^{(3)}(\mathbf{p} + \mathbf{p}_\pi - \mathbf{k}-\mathbf{q})\nonumber\\
& \times \delta(\omega-E+m_N-e({\mathbf k}+{\mathbf q}-{\mathbf p}_\pi) - e_\pi(\mathbf{p_\pi}))\, ,
\label{had:tens_pi}
\end{align}
where $e_\pi(\mathbf{p_\pi})=\sqrt{{\mathbf p}^2 + m_\pi^2}$ is the energy of the outgoing pion and $t_\pi$ its isospin. Again, the phase space integrals over the momenta and the spin-isospin summations of the final-state particles are performed outside the hadron tensor. The main difference between the above expression and \cref{had:tens} resides in the elementary amplitude. To describe the pion-production processes, we need matrix elements of the
charged-current operator causing the transition from a bound
nucleon $|k \rangle$ to a state with a pion and a nucleon
$|p_\pi p\rangle$.

The present work uses the ANL-Osaka DCC model discussed above to generate the required matrix elements of the charged-current operator describing the electroweak transition from a bound nucleon state $|k\rangle$ to a pion nucleon state $|p_\pi p\rangle$.
The necessary matrix elements are obtained from tables of amplitudes in the helicity - $LSJ$ mixed representation of Ref.~\cite{Kamano:2013iva}. This corresponds to a partial wave expansion of the final-state pion nucleon pair, as in Appendix~\ref{app:MB_amplitudes}, while the initial state is described by two-particle helicity states with total angular momentum $J$.
From these amplitudes the matrix elements $\langle p_\pi, p; s_p  \rvert j^\mu \rvert k ; s_p \rangle$, with $s_p(s_k)$ the final(initial) nucleons spin, are computed in the center of momentum system. They are related to the matrix elements of Eq.~(\ref{had:tens_pi}) by a boost as described in Ref.~\cite{Rocco:2019gfb}. In particular, the Wigner rotation of the spins is properly taken into account, thereby all angular dependence of the matrix element is included.
The phase space is again handled using standard techniques developed for the LHC (see App.~\ref{app:phase_space}).
The results obtained neglecting FSI have been validated against inclusive electron scattering data on nuclei in Ref.~\cite{Rocco:2018mwt,Rocco:2019gfb} and have been interfaced to \achilles using the \texttt{Fortran90} wrapper discussed in Ref.~\cite{Isaacson:2022cwh}.

\subsection{Intranuclear Cascade}
The \achilles INC begins by randomly sampling spatial distributions of protons and neutrons. For the $^{12}$C nucleus, these are obtained from Green’s function Monte Carlo calculations~\cite{Carlson:2014vla}, thereby fully retaining correlation effects~\cite{Isaacson:2020wlx}. On the other hand, despite recent progress in utilizing artificial neural networks to compactly represent nuclear many-body wave functions~\cite{Gnech:2023prs}, $^{40}$Ar is not yet amenable to continuum quantum Monte Carlo methods. In this case, the spatial positions of protons and neutrons are sampled from the single-proton and single-neutron density distributions, neglecting correlation effects altogether~\cite{Isaacson:2020wlx}.
Particles produced from hard scattering events are propagated through the nucleus in straight-line trajectories in small steps of time $\delta t$.\footnote{The option to bend these trajectories using nuclear potentials is available in \achilles, see Ref.~\cite{Isaacson:2022cwh} for details.}
At every time step, propagating particles have the ability to interact with spectator nucleons based on their impact parameter and the available cross sections.
The probability of an interaction occurring is based on a Gaussian or cylinder probability model which depends on both the impact parameter and the total cross section.
In Refs.~\cite{Isaacson:2020wlx,Isaacson:2022cwh} this approach was validated based on hadron-nucleus and electron-nucleus scattering data. Below we describe newly incorporated hadron-nucleon scattering models needed to accommodate pion propagation in the nucleus. 

\subsubsection{ANL-Osaka DCC Model}
To incorporate meson-baryon interactions in the INC, we make use of the standard decomposition of the scattering matrix into partial-wave amplitudes (PWAs).
For completeness, definitions of the PWA are collected in App.~\ref{app:MB_amplitudes}.
In addition to $\pi N$, the other zero strangeness meson-baryon channels ($\eta N$, $K\Lambda$, $K\Sigma$) are included.
For consistency with the pion production model, the amplitudes are all obtained from the ANL-Osaka DCC model.

The total meson-baryon cross section is the sum of the angle-integrated cross sections for all kinematically accessible channels.
The cross section for each process $i \rightarrow f$, where $i,f$ denote a meson-baryon state is given in terms of squared PWAs
\begin{equation}
\sigma_{if}(s) = \frac{4\pi}{\lvert \mathbf{k}_i \rvert^2} \sum_L \left( L\lvert \tau^{L,-}_{if}(s)\rvert^2 + (L+1) \lvert \tau^{L,+}_{if}(s)\rvert^2\right),
\end{equation}
where $\sqrt{s}$ is the center-of-mass energy, and $\mathbf{k}_i$ the center-of-mass momentum of the initial pair. The reduced PWA for given orbital angular momentum $L$, and total angular momentum $J=L\pm 1/2$, $\tau^{L,\pm}_{if}$, are defined in App.~\ref{app:MB_amplitudes}.
A particular meson-baryon scattering event is selected according to the weight given by  the partial cross sections $\sigma_{if}(s)$.
The final-state kinematics are then determined by the cross section in the center-of-momentum frame.
The CM scattering angle $z= \hat{\mathbf{k}}_i\cdot\hat{\mathbf{k}}_f = \cos\theta$ of the meson-baryon pair is generated according to the cross section 
\begin{align}
\label{eq:sigma_dif}
    \frac{\mathrm{d}\sigma_{if}}{\mathrm{d}\Omega} = &\frac{1}{\lvert \mathbf{k}_i \rvert^2} \Big\lvert \sum_L P_L(z) \left[ L~\tau^{L, -}_{if}(s) + (L+1)~\tau^{L,+}_{if}(s)\right] \Big\rvert^2 \nonumber\\
       &+ \frac{1-z^2}{\lvert\mathbf{k}_i\rvert^2} \Big\lvert \sum_L~ P_L^\prime(z) \left[\tau^{L,+}_{if}(s) - \tau^{L,-}_{if}(s)\right] \Big\rvert^2,
\end{align}
where $P_L$ and $P'_L$ denote the Legendre polynomials of order $L$ and their derivatives. Validations of both the total and angular differential cross sections can be found in App.~\ref{app:MBvalidation}.
The distribution of the azimuthal angle in the CM frame is uniform.

While the interference between different partial waves cancels for the angle-integrated cross section, the angular distribution does include interference terms.
We use the full angular distribution given by the polynomial of Eq.~(\ref{eq:sigma_dif}), including all interferences exactly.
Since the PWAs constitute a model-independent, complete basis for the scattering matrix, any approach for meson-baryon scattering in the resonance region, such as  the J\"ulich-Bonn-Washington analysis~\cite{Ronchen:2012eg}, can be included straightforwardly.

Note that we choose to include this set of two-body channels explicitly, but the full DCC analysis includes additional final states, notably the $\rho N, \sigma N$, and $\pi \Delta$ channels which are doorway states for two-pion production.
This subset of channels will underpredict the total cross section obtained from the optical theorem 
\begin{equation}
    \sigma_{i\rightarrow X} = \frac{2\pi}{\lvert \mathbf{k}_i \rvert^2} \sum_{L,J} (2J+1) \mathrm{Im}\left[\tau^{L,J}_{ii} \right].
\end{equation}
It is possible to rescale the total cross section based on this value, but to saturate this total rate with the correct particle content, one needs to include explicitly multi-pion final-states. This will be pursued in future work. 

\subsubsection{Oset model}
Another common approach to pion-nucleon interactions in neutrino event generators involves the optical-potential model of Oset~\cite{Salcedo:1987md}. The model begins with the pion-nucleus optical potential, which is computed in infinite nuclear matter and corrected for finite nuclei by using the local density approximation. The model also includes certain in-medium effects such as Pauli-blocking and in-medium modifications to the $\Delta$ propagator. The p-wave probability for a pion to interact in the nuclear medium is obtained by computing the $\Delta$ self energy associated with the excitation of a $\Delta$-hole pair. Interactions in the nuclear medium modify the $\Delta$, giving rise to a self-energy function with imaginary part
\begin{align}
    \begin{split}
    {\rm{Im}}\Sigma_{\Delta}(T_{\pi})
    &= -[C_{Q}(\rho/\rho_{0})^{\alpha}\\
    &+ C_{A2}(\rho/\rho_{0})^{\beta} + C_{A3}(\rho/\rho_{0})^{\gamma}]\, ,
    \end{split}
    \label{eq:delta_self_energy}
\end{align}
where $\rho_{0} =0.16/\rm{fm}^{3}$ is the nuclear saturation density and $\rho$ is the local nuclear density.

The different terms in \cref{eq:delta_self_energy} arise from quasi-elastic scattering, and pion absorption on two and three nucleons respectively. The coefficient functions $C_{Q}, C_{A2}, C_{A3}$, and the corresponding exponents are parameterized in Ref.~\cite{Oset:1987re} in terms of the pion kinetic energy, $T_{\pi}$, up to 350 MeV. In this work evaluate \cref{eq:delta_self_energy} for pion kinetic energies beyond this value to compute absorption cross sections and comment on this effect in Sections~\ref{sec:validation} and~\ref{sec:compare_e_nu}.
Oset \emph{et al}. also provide a parameterization of the s-wave, non-resonant part of the optical potential which is important at $T_{\pi} < 50$ MeV. 

Though this model can be used to compute microscopic cross sections for $\pi N$ quasi-elastic scattering, we utilize it only to compute pion absorption cross sections as follows. Consider first the case of absorption on two nucleons. When a candidate pion-nucleon pair is chosen for possible interaction, the absorption cross section is computed by searching for a nearby secondary nucleon to participate in the interaction. This search involves all isospin combinations, i.e., if the interacting pair is a $\pi^{+}n$ pair, then both a secondary neutron for $\pi^{+}nn\rightarrow pn$ and a secondary proton for $\pi^{+}np\rightarrow pp$ are considered.
To select the actual interaction experienced by the candidate pair,
the cross sections are weighted according to the isospin dependence of the elementary cross sections~\cite{VicenteVacas:1993bk}.
In this way the absorption cross sections on asymmetric nuclei can be consistently computed as soon as separate proton and neutron densities are given.
For simplicity we have only implemented the two nucleon absorption process, though we retain the three-nucleon absorption term to maintain the overall rate.
We leave the inclusion of genuine 4-body initial states to future work.

\subsubsection{GiBUU Delta model}
An alternate approach to pion-nucleus interactions, used in the GiBUU event generator~\cite{Buss:2011mx} and the Li\'ege INC~\cite{Cugnon:1996kh}, treats the $\Delta$ as a propagating degree of freedom.
Although explicit propagation of unstable particles introduces certain complications (namely, requiring models for their production, interactions, and  decay), its utility extends beyond the simple case of the $\Delta$ resonance. 
For example, a framework for propagation of unstable particles provides the ability to decay final-state particles such as $\pi^0\to\gamma\gamma$.

When the $\Delta$ is a propagating degree of freedom, pion interactions occurs via a two-step process, where first a $\pi N$ interaction creates a $\Delta$ which propagates through the nucleus, and second this $\Delta$ either re-interacts or decays back to a $\pi N$. For example, pion absorption would occur via
\begin{align}
    \pi N &\rightarrow \Delta \\
    \Delta N &\rightarrow N N.
\end{align}
Decays are handled probabilistically at each time step according to $P(t) =1 -  e^{-\delta t/\tau}$, where $\delta t$ is the time step for the cascade, and $\tau=\gamma\hbar c/\Gamma$ is the lifetime in the lab frame of the particle with a vacuum width of $\Gamma$.
When a decay occurs, the decay channel is selected according to the branching ratios.\footnote{Currently, only two-body decays are implemented in \achilles.}
The decay is rejected if any of the decay products are Pauli blocked, thereby capturing some in-medium effects on the resonance width.

Additional interaction channels (like $N N \rightarrow N \Delta$ and $N\Delta \rightarrow N\Delta$) are physically possible and indeed required for theoretical consistency.
Detailed balance~\cite{landau1981physical} relates the reactions $N N \rightarrow N\Delta$ and $N\Delta\rightarrow NN$:
\begin{equation}
    \sigma_{cd\rightarrow ab} = \sigma_{ab\rightarrow cd} \left(\frac{p_{ab}}{p_{cd}}\right)^2 \frac{(2J_a+1)(2J_b+1)}{(2J_c+1)(2J_d+1)}\frac{\mathcal{S}_{ab}}{\mathcal{S}_{cd}}\,,
    \label{eq:detail_balance}
\end{equation}
where $p_{ij}=|\mathbf{p}_i|=|\mathbf{p}_j|$ is the center-of-mass three-momentum of particles $i$ or $j$, $J_i$ is the total spin of particle $i$, and $\mathcal{S}_{ij}$ is a symmetry factor 
for particles $i$ and $j$, e.g., $1/2$ for identical particles.

The cross section for the two-body $NN\leftrightarrow N\Delta$ scattering process can be generically parameterized as
\begin{equation}
    \frac{\rm{d}\sigma_{NN\rightarrow N\Delta}}{\rm{d}\mu^2_{\Delta}\rm{d}\Omega} = \frac{\lvert\mathcal{M}_{NN\rightarrow N\Delta}\rvert^2}{64\pi^2s}\frac{p_{N\Delta}}{p_{NN}}\mathcal{A}_\Delta(\mu^2_\Delta)\,.
    \label{eq:nntondelta}
\end{equation}
The resonance spectral function $\mathcal{A}_\Delta(\mu_\Delta^2)$ for the $\Delta$ is related to the Breit--Wigner propagator and is given by
\begin{equation}
    \mathcal{A}_\Delta(\mu_\Delta^2) = \frac{1}{\pi}\frac{\mu_\Delta \Gamma}{(m_\Delta^2-\mu_\Delta^2)^2+\mu_\Delta^2\Gamma^2}\,,
\end{equation}
where $m_\Delta$ is the on-shell mass, $\mu_\Delta$ is the off-shell mass, and $\Gamma = 112$~MeV is the width.
Ref.~\cite{Dmitriev:1986st} 
(see also recent theses~\cite{Buss:2008zz, Effenberger:1999wlg})
has computed the process $pp\rightarrow n\Delta^{++}$ at leading order using the interaction Lagrangian
\begin{align}
\begin{split}
    \mathcal{L} = \frac{F(t)}{m_\pi}\Big[& f_{NN\pi} (\bar{N}\gamma_\mu\gamma_5 \tau_a N) \partial^\mu \Pi_a \\
    & + f_{N\Delta\pi} \sqrt{z}\, (\bar{\Delta}_\mu T_a N) \partial^\mu \Pi_a\Big],
\end{split}
\label{eq:DeltaProductionLagrangian}
\end{align}
where $\Delta_\mu$ is a spin-$\tfrac{3}{2}$ Rarita--Schwinger field for the $\Delta$-baryon, $N$ is a spin-$\tfrac{1}{2}$ field for the nucleon, and $\Pi_a$ is the pion field.
Where $\tau_a$ are the Pauli isospin matrices describing the $\tfrac{1}{2}\to \tfrac{1}{2}$ transition while $T_a$ represent the 2$\times$4 isospin coupling matrices which obey the relation 
\begin{equation}
T_a T_b^\dagger = \mathbb{I}\,\delta_{ab} -\frac{1}{3}\tau_a \tau_b
\end{equation}
for the $\tfrac{1}{2}\to\tfrac{3}{2}$ processes.
The dimensionless coefficients $f_{NN\pi}=1.008$ and $f_{N\Delta\pi}=2.202$ are the effective couplings of the pion to the nucleon and $\Delta$, respectively.\footnote{These numbers differ slightly from those of the DCC model but are consistent with Ref.~\cite{Dmitriev:1986st}.}
The vertex form factor $F(t)$ is taken to have the form
\begin{equation}
    F(t)=\frac{\Lambda^2-m_\pi^2}{\Lambda^2-t}\,,
    \label{eq:delta_formfactor}
\end{equation}
with $\Lambda=0.63$ GeV~\cite{Dmitriev:1986st}.
Each $\Delta N \pi$ vertex also gives an additional form factor $\sqrt{z}$ of the form 
\begin{equation}
    z(t,\mu_\Delta)=\frac{\lambda(\mu_\Delta,m_N,\sqrt{t})+\kappa^2}{\lambda(m_\Delta,m_N,\sqrt{t})+\kappa^2}\,,
\end{equation}
where $\lambda(a,b,c) = a^2 + b^2 + c^2 - 2ab-2ac-2bc$ is the K\"all\'en function,
and $\kappa=0.2$ GeV is a parameter obtained from a fit to $\pi^+ p\to\pi^+ p$ scattering data for $\mu_\Delta<1.4$~GeV~\cite{Dmitriev:1986st}.
Additional details on the calculation of the matrix elements and a validation of the $NN\rightarrow  NN\pi$ cross section are given in App.~\ref{app:delta_mat}.

In addition to $\Delta$ production, we also include elastic scattering of $\Delta$'s with nucleons (\textit{i.e.} $N\Delta\rightarrow N\Delta$).
The implementation in \achilles follows the calculation done in Refs.~\cite{Effenberger:1999wlg,Buss:2008zz,Buss:2011mx}.
The differential cross section is given as
\begin{equation}
    \frac{{\rm d}^2\sigma}{{\rm d}\mu_{\Delta_f}{\rm d}\Omega} = \frac{|\mathcal{M}_{N\Delta_i\to N\Delta_f}|^2}{64\pi^2s}\frac{p_{N\Delta_f}}{p_{N\Delta_i}}\mathcal{A}_\Delta(\mu_f^2),
\end{equation}
where $p_{N\Delta_i}$ and $p_{N\Delta_f}$ are the center-of-mass momentum of the initial and final states respectively, and $\mu_{\Delta_f}$ is the off-shell mass of the final state $\Delta$.
The matrix element is again obtained from the one-pion exchange model. 
See App.~\ref{app:delta_mat} for further details.

Following the work of Ref.~\cite{Bogart:2024gmb} where it was shown that in-medium changes to the $NN\to N\Delta$ cross sections improve agreement with neutrino scattering data on ${}^{40}\rm{Ar}$, we implement as an option the density suppression of this cross section following the work of Song and Ko~\cite{Song:2014xza}. This modifies the resonance excitation in $NN$ collisions according to
\begin{equation}\label{eq:density_sup}
    \sigma_{NN\to N\Delta}(\rho) = \rm{exp}\left(-\alpha \frac{\rho}{\rho_{0}}\right)\sigma_{NN\to N\Delta}(0)\, ,
\end{equation}
where $\alpha$ is a tunable parameter and $\rho$ is the local nuclear density. This affects both pion production via \cref{eq:density_sup} as well as pion absorption through the reversed process $N\Delta\to NN$. For reference, the authors of Ref.~\cite{Bogart:2024gmb} use a value of $\alpha = 1.2$. We found that including this effect reduces overall agreement with experimental data, and therefore set $\alpha=0.0$ for the results shown in this paper.

\begin{table}[t!]
\centering
\caption{
Available initial and corresponding final states in the Virtual Resonances and Propagating Resonances modes of the \achilles cascade. 
The final row with initial state $\pi NN$ corresponds to pion absorption; the corresponding $s$- or $p$-wave contributions are computed using the Oset model.
The $s$-wave absorption in the Propagating Resonances mode also uses the Oset model to capture this process (see the text for more details).
As indicated, not all initial states are present in each case.
\label{tab:cascade run modes}
}
\begin{tabular}{c|c|c}
\hline\hline
Initial & \multicolumn{2}{c}{Final states in different cascade modes} \\
\cline{2-3}
States& Virt. Resonances mode & Prop. Resonances mode \\
\hline
$NN$ & $NN, NN\pi$ & $NN, N\Delta(\to NN\pi, NN\gamma)$\\
$\pi N$ & $\pi N, \eta N, K \Lambda, K \Sigma$ & $\Delta (\to \pi N, \gamma N)$ \\
$\eta N$ & $\pi N, \eta N, K \Lambda, K \Sigma$ & Initial state not present \\
$\Delta N$ & Initial state not present & $NN, \Delta N$ \\
$\pi N N$ & $NN$ ($s$- and $p$-wave) & $\Delta N(\to NN)$, $NN$ ($s$-wave) \\
\hline\hline
\end{tabular}
\end{table}

Finally, in addition to allowing the $\Delta$ to propagate in $NN\to N\Delta\rightarrow NN\pi$, we offer the option to immediately decay the $\Delta$. This allows us to compute the $\Delta$ resonance contribution to $NN\to NN\pi$ in models where the $\Delta$ is not a propagating degree of freedom.

\subsubsection{\achilles Cascade Modes}
The above models contain all the ingredients necessary for pion propagation within the nucleus. In contrast to the propagating resonances which mediate $\pi N$ interactions in the GiBUU $\Delta$ model, the $\pi N$ interactions in the Oset and ANL-Osaka DCC model in the Virtual Resonances cascade are time independent point interactions (i.e. the production and decay vertices are not displaced).
This is due to the fact that the {\it virtual} resonances are already integrated out when computing the interaction cross sections. 

Given this, we have developed two cascade modes dubbed {\it Virtual Resonances} and {\it Propagating Resonances}. Table~\ref{tab:cascade run modes} summarizes the ingredients of these two modes.

The Virtual Resonances mode relies on the ANL-Osaka DCC model to simulate meson-baryon to meson-baryon scattering (i.e. $\pi N\rightarrow\pi N$), the Oset model for pion absorption, and utilizes the $\Delta$ resonance contribution of the $NN\to NN\pi$ process where the intermediate $\Delta$ is immediately decayed. The Propagating Resonances mode mainly uses the GiBUU $\Delta$ model with a propagating $\Delta$, but also includes the S-wave piece of the Oset model for pion absorption. This is included to reproduce the behavior of the pion absorption cross section at lower pion kinetic energies. The Propagating Resonance mode neglects interference effects and higher mass resonances between different interaction channels in the $\pi N \to \pi N$ chain that are included within the ANL-Osaka DCC model. In future work, we plan to include the higher resonances to fully handle the ANL-Osaka DCC model range of validity.
The difference between these two choices gives an estimate of the uncertainty associated with modeling resonances during the intranuclear cascade.

\begin{figure*}[t!]
    \includegraphics[width=\textwidth]{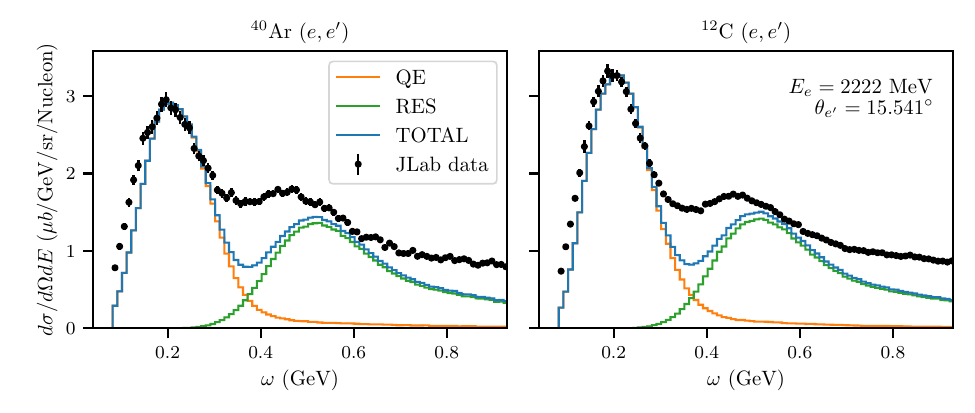}
    \caption{\achilles comparison to inclusive electron scattering data on ${}^{40}\rm{Ar}$ (left) and ${}^{12}\rm{C}$ (right) from JLab at $E_{\rm{beam}} = 2.222$ GeV and $\theta_{e^{\prime}} = 15.541^\circ$~\cite{Murphy:2019wed} as a function of energy transfer ($\omega$). The data and predictions have been normalized to the number of nucleons in each respective nucleus. The total prediction from \achilles is in blue, while the quasi-elastic component is in orange and the pion production component is in green.
    Gaps between the data and \achilles' prediction for the differential cross section are expected to be largely filled by the eventual inclusion of contributions from meson exchange currents and deep inelastic scattering.
    }
    \label{fig:inclusive_data}
\end{figure*}

\section{Validation}
\label{sec:validation}
\subsection{Hard Scattering}

The hard scattering model of Sec.~\ref{sec:theory} was validated against $(e,e^\prime)$ data on ${}^{12}\rm{C}$ at several beam energies and scattering angles in Ref.~\cite{Rocco:2019gfb}.
For these comparisons the authors of Ref.~\cite{Rocco:2019gfb} used the spectral function of ${}^{12}\rm{C}$ computed using correlated-basis function theory~\cite{Benhar:1994hw}, which is also used in \achilles~\cite{Isaacson:2022cwh}.
Recently, high-precision $(e,e^\prime)$ and $(e,e^\prime p)$ data was recorded at Thomas Jefferson Lab (JLab) on ${}^{12}\rm{C}$, ${}^{40}\rm{Ar}$, and ${}^{48}\rm{Ti}$~\cite{Murphy:2019wed,JeffersonLabHallA:2022cit,JeffersonLabHallA:2022ljj}. These experiments were intended to test the scaling behavior of the $(e,e^\prime)$ cross section and to study nuclear effects in $(e,e^{\prime}p)$ scattering on ${}^{40}\rm{Ar}$ and ${}^{48}\rm{Ti}$, relevant for the next generation of liquid-argon based neutrino experiments.
To compare to data taken on ${}^{40}\rm{Ar}$, we have incorporated the proton and neutron spectral function of ${}^{40}\rm{Ar}$ extracted from $(e,e^\prime p)$ data on ${}^{40}\rm{Ar}$ and ${}^{48}\rm{Ti}$~\cite{JeffersonLabHallA:2022cit,JeffersonLabHallA:2022ljj,Nikolakopoulos:2024mjj}. 
Figure~\ref{fig:inclusive_data} shows the \achilles comparison to a selection of $(e,e^\prime$) data taken at $E_{\rm{beam}} = 2.222$ GeV and $\theta_{e^\prime} = 15.541^\circ$ on ${}^{12}\rm{C}$ and ${}^{40}\rm{Ar}$ including contributions from quasi-elastic scattering and single-pion production.

\begin{figure}[t!]
    \centering
    \includegraphics[width=0.5\textwidth]{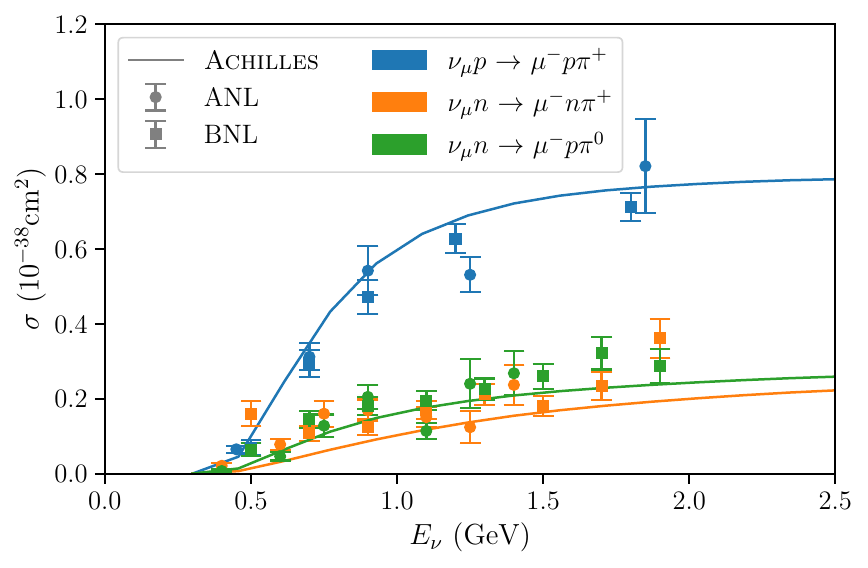}
    \caption{Comparison of total pion-production cross sections for $\nu_\mu$ scattering on nucleons. 
    The data are from measurements at Argonne and Brookhaven National Laboratories reanalyzed in Ref.~\cite{Wilkinson:2014yfa}.
    }
    \label{fig:anl_bnl}
\end{figure}

The quasi-elastic (orange) and $\Delta$ peaks (green) in both nuclei are qualitatively reproduced.
The significant difference between the data and predictions in the dip region between the QE and $\Delta$ peaks is due to the lack of meson-exchange currents (MEC) in \achilles, known to fill this region~\cite{Rocco:2019gfb}. 
Similarly, the deficit at high energy transfers beyond the $\Delta$ peak can be explained by the lack of multi-pion production and DIS in \achilles. 
The inclusion of these additional hard scattering processes will be the focus of future efforts. 
The small shift in the QE peak relative to the data is due to neglected final-state interactions in the hard-scattering matrix elements. 
The effect of final-state interactions is twofold: the dispersion relation of outgoing hadrons is modified\footnote{The full final-state system is an eigenstate of the interacting many-body Hamiltonian.
The momentum of any single nucleon is not a conserved quantum number, essentially the nucleon and residual system may exchange momentum, and $|p\rangle \otimes |\Psi_f^{A-1}\rangle$ only asymptotically resembles a free nucleon with fixed momentum and a $(A-1)$ system.}, and strength is redistributed over asymptotic final-state channels.
The INC takes care of the latter; crucially it goes beyond the optical-potential approach by explicitly populating accessible final-states~\cite{Nikolakopoulos:2024mjj, Nikolakopoulos:2022qkq}. 
The former affects the nuclear response integrated over final-states, and hence the hard scattering matrix element of Eq.~(\ref{eq:genfact}). This can be incorporated within the factorized picture via a folding function approach~\cite{Benhar:2006hh, Rocco:2019gfb}, by treating the outgoing particles as scattering states in a suitable real potential~\cite{, Gonzalez-Jimenez:2019qhq} or through the (relativistic) Green's function approach of Refs.~\cite{Meucci:2003cv,PhysRevC.22.1680,Meucci:2003uy}. The effect in this kinematic region is a slight shift toward lower energy transfers and reduction of the height of the quasi-elastic peak~\cite{Gonzalez-Jimenez:2019ejf}. Implementing this in a manner that is consistent with the intranuclear cascade exceeds the scope of the present work.
We also compare the \achilles predictions for the total cross section for pion production on nucleons against the reanalyzed bubble chamber data from ANL and BNL~\cite{Wilkinson:2014yfa}. These include $\nu_{\mu}p\to\mu^{-}p\pi^{+}$, $\nu_{\mu}n\to\mu^{-}n\pi^{+}$, and $\nu_{\mu}n\to\mu^{-}p\pi^{0}$, the latter being extracted from data on deuterium instead of free neutrons. 
Figure \ref{fig:anl_bnl} shows that the implementation of the ANL-Osaka DCC model in \achilles reproduces the neutrino energy dependence of these elementary cross sections very well. 

\subsection{Pion-Nucleus scattering}
\begin{figure*}[t!]
    \centering
    \includegraphics[width=1.0\textwidth]{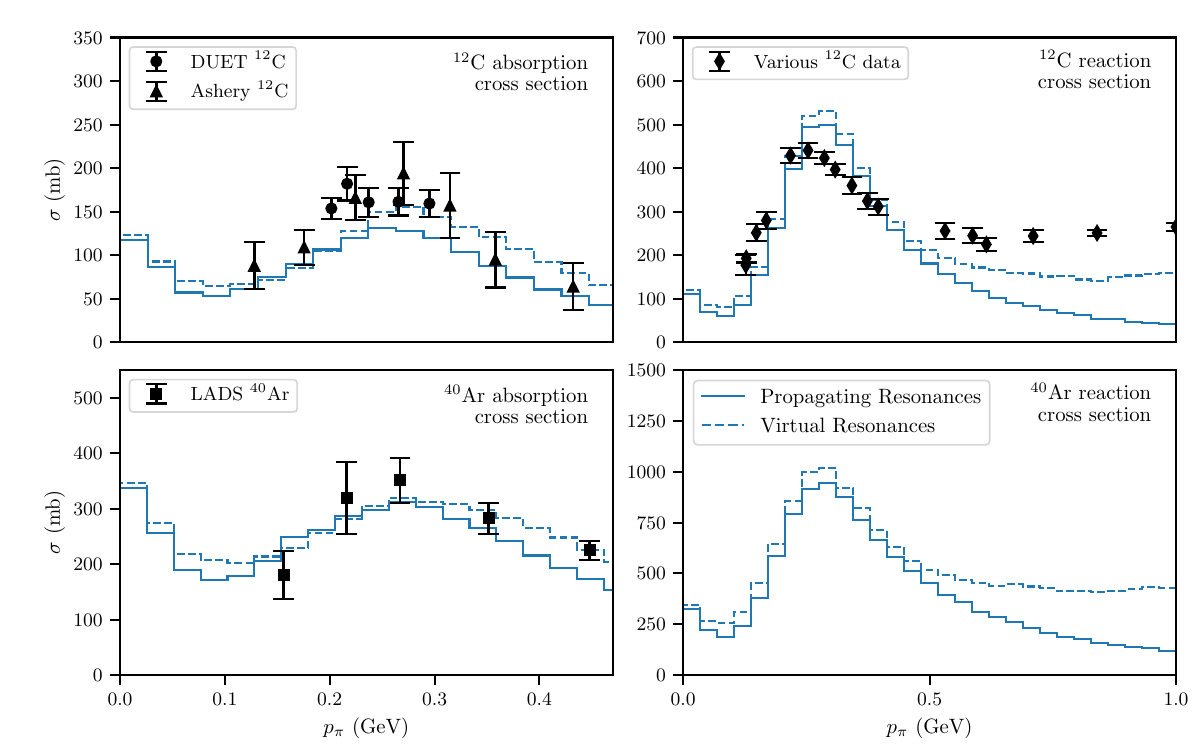}
    \caption{
    Validation of the pion modes in \achilles INC against experimental data. \achilles results are shown using both the 
    Propagating Resonances model (solid line) and the Virtual Resonances model (dotted line).
    \textbf{Left:}
    Absorption cross section for $\pi^{+}$ on carbon (top) and $\pi^{+}$ on argon (bottom). 
    Experimental data are for $\pi^+$ absorption measurements on carbon (DUET~\cite{DUET} and Ashery~\cite{ASHERY}) and on argon (LADS~\cite{LADS:2000pzh}).
    \textbf{Right:}
    Total reaction cross section for $\pi^{+}$ on carbon (top) and $\pi^{+}$ on argon (bottom).
    Data for carbon are from Ref.~\cite{Dytman:2021ohr}.
    }
    \label{fig:c12_ar40_validation}
\end{figure*}

We compute the $\pi^{+}$-carbon, and $\pi^{+}$-argon reaction and absorption cross sections. The results for the reaction cross sections, using both the Virtual and Propagating Resonances model are shown in Fig.~\ref{fig:c12_ar40_validation}. We remind the reader that the DCC piece of the Virtual Resonances model uses only the contribution from explicit final-state meson-baryon channels $(\pi N, \eta N, K \Lambda, K \Sigma)$ without in-medium modification of the cross sections.
Both models undershoot the data for the reaction cross section on carbon in the high-$p_\pi$ tail. This is to be expected since two-pion production and other channels are not included.
Surprisingly, both model predictions exceed the data near the $\Delta$ peak, around $p_{\pi}=300$ MeV, which could point to the need to incorporate in-medium modifications of the $\pi N$ cross sections in both models.
Other explanations are also possible. For instance, the present treatment describes the nucleon momentum distribution using a local Fermi gas. A more realistic distribution would include some high-momentum nucleons, which would lead to additional smearing of features. 

Figure~\ref{fig:c12_ar40_validation} compares the predictions of the absorption cross section to $\pi^{+}$ data on carbon and argon. 
Both models slightly underestimate the data for the absorption cross section on carbon at pion momenta around the $\Delta$ peak. 
For argon, both models give excellent agreement with the LADS data~\cite{LADS:2000pzh}, except for the Propagating Resonances model at the highest pion momentum. We note that the direct $NN\Delta\to NNN$ process is missing in the Propagating Resonance model which when included could improve agreement with the data. At the moment the \achilles cascade does not handle three-body initial states. Additionally, we note that the strength at low pion momentum comes almost entirely from the Oset S-wave absorption process. Although experiments on the deuteron have shown this to be an important process at low energies~\cite{Vogelzang:1985yn}, little is known about this process in heavier nuclei. We leave the investigation of these two aspects for future work.

As mentioned previously, we have extend the use of the Oset Model beyond its original limit of $T_{\pi} = 350$ MeV in the Virtual Resonances cascade. We have investigated the effect of this by freezing the various pion absorption cross sections at kinetic energies beyond this limit to their values at $T_{\pi} = 350$ MeV. As expected, this only affects the $\pi$-carbon and $\pi$-argon absorption cross sections above $T_{\pi} = 350\,\rm{MeV}\,(p_{\pi}\approx 470\,\rm{MeV})$, where the absorption cross sections are reduced by roughly 30\%. The agreement with the data is not qualitatively changed. We have confirmed that this change has very little effect on the various electron- and neutrino-nucleus observables we will compare to, and we leave it to further work to analyze this approximation for higher energy experiments.

\section{Exclusive electron- and neutrino-nucleus scattering}
\label{sec:compare_e_nu}
Modern accelerator neutrino experiments utilize broad-band beams spanning the energy range from hundreds of MeV to 10s of GeV. 
This enables us to test neutrino-nucleus interaction models against multiple nuclear targets over a wide range of energies where multiple processes contribute to the observed event rate. Additionally, electron-nucleus scattering has become invaluable for constraining the vector part of the nuclear electroweak current~\cite{Ankowski:2022thw,e4vincl,CLAS:2021neh}. Electrons probe the same nuclear ground state as neutrinos, and hadrons leaving the nucleus undergo similar final-state interactions. 
In this section we update the comparisons made in Ref.~\cite{Isaacson:2022cwh} with exclusive electron-carbon scattering data from the e4$\nu$ collaboration~\cite{CLAS:2021neh}, as well as present comparisons with flux-folded neutrino scattering data from T2K, MINER$\nu$A and MicroBooNE. 
These experiments operate in complementary beam lines and probe targets composed of hydrocarbon (MINER$\nu$A and T2K), and argon (MicroBooNE). 
A global comparison including these experiments highlights the need to capture both the energy and $A$ dependence of neutrino-nucleus cross sections.

\subsection{Comparison to e4$\nu$ \label{sec:e4v}}
The CLAS and e4$\nu$ collaborations have analyzed data on electron-nucleus scattering at similar energies- and using similar hadron detection thresholds as neutrino scattering experiments~\cite{CLAS:2021neh}. This analysis focused on the comparison of different neutrino energy estimators experiments used compared to the known electron beam energy. Although data for carbon, iron, and helium-4 are available at 1.159, 2.257, and 4.453 GeV, we focus on comparisons with carbon using the lowest electron beam energy, 1.159 GeV, as this energy range is the least impacted by DIS and multi-pion production. The choice of carbon is to compare against neutrino experiments utilizing similar targets.

\begin{figure}
    \centering
    \includegraphics[width=1.0\linewidth]{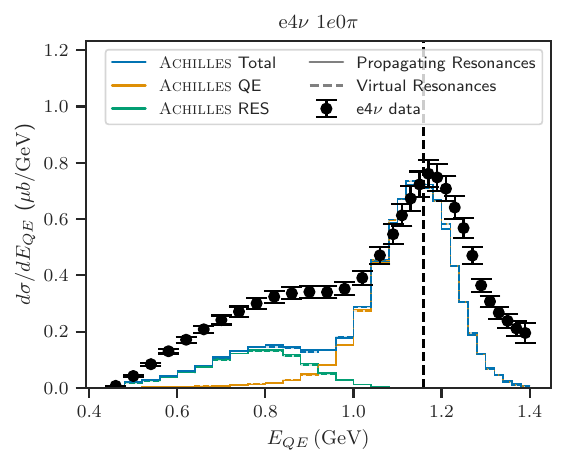}
    \caption{
    Comparison of the electron-carbon $0\pi$ differential cross section with respect to $E_{QE}$ measured by e4$\nu$ and CLAS~\cite{CLAS:2021neh} compared to predictions by \achilles.
    The black vertical dashed line shows the true incoming beam energy.}
    \label{fig:e4v0pi}
\end{figure}

Figure \ref{fig:e4v0pi} shows comparisons to the $0\pi$ cross section measurement by e4$\nu$ differential in $E_{\rm QE}$, defined as
\begin{equation}\label{eq:EQE}
    E_{\rm QE} = \frac{2m_{N}\epsilon + 2m_{N}E_{e'} - m_{e'}^2}{2(m_{N} - E_{e'} + p_{e'}\cos\theta_{e'})}\,,
\end{equation}
where the average separation energy $\epsilon = 21~\mathrm{MeV}$ was determined in the experimental analysis.
This quantity is an energy estimator used by Cherenkov detectors like Super-Kamiokande and MiniBooNE where outgoing hadrons are often under detection threshold, and each neutrino event is assumed to be produced via quasi-elastic neutrino-nucleus scattering. The only relevant measured kinematic variables are the outgoing lepton energy and scattering angle. 
Events with $E_{\rm QE}$ reconstructing near the incoming beam energy are dominated by quasi-elastic scattering, whereas those in which a pion was produced and then absorbed during the cascade or was under threshold for detection reconstruct to lower values of $E_{\rm QE}$. We note that the deficit of events in \achilles at large $E_{\rm QE}$ might be partially caused by the neglect of interference effects in the cascade and the effect of final-state interactions on the hard scattering matrix element which lead to a shift to lower energy transfer. The latter cause a modest shift in the peak, and MEC will also contribute a tail at large $E_{QE}$~\cite{Nikolakopoulos:2018sbo}. The deficit at low $E_{\rm QE}$ will be improved once MEC and DIS are incorporated. Both cascade modes give similar results in the region dominated by resonance production, meaning that pions are absorbed at similar rates in both models at this beam energy. 

\begin{figure}
    \centering
    \includegraphics[width=1.0\linewidth]{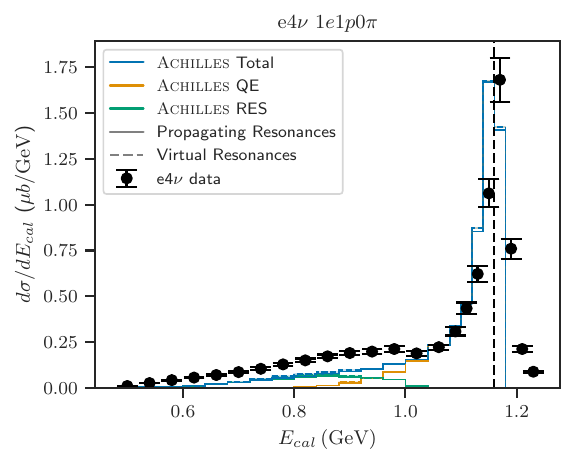}
    \caption{Comparison of the electron-carbon $1p0\pi$ differential cross section with respect to $E_{cal}$ measured by e4$\nu$ and CLAS~\cite{CLAS:2021neh} compared to predictions by \achilles. The black vertical dashed line shows the true incoming beam energy.}
    \label{fig:e4v1p0pi}
\end{figure}

\begin{figure}
    \centering
    \includegraphics[width=1.0\linewidth]{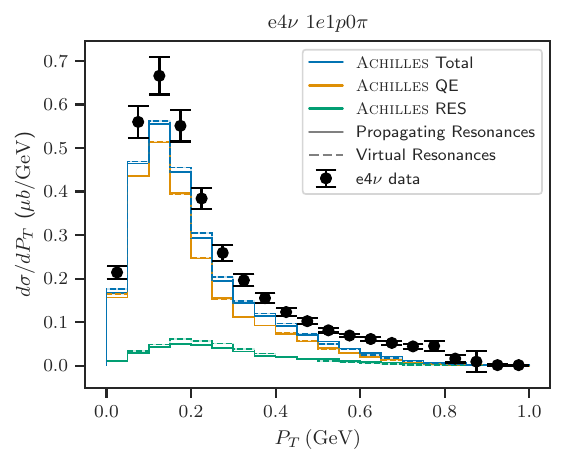}
    \caption{Comparison of the electron-carbon $1p0\pi$ differential cross section with respect to $P_{T}$ measured by e4$\nu$ and CLAS~\cite{CLAS:2021neh} compared to predictions by \achilles.
    }
    \label{fig:e4v1p0piPT}
\end{figure}

Moving to more exclusive final states, Figs.~\ref{fig:e4v1p0pi} and~\ref{fig:e4v1p0piPT} show comparisons to the $1p0\pi$ cross sections differential in $E_{\rm cal}$ as well as $P_{T}$. 
This more restrictive signal reduces the contribution from resonance and DIS events. The calorimetric reconstructed energy is the neutrino energy estimator used by detectors like MINER$\nu$A, MicroBooNE, and the future DUNE experiment, where the energies of outgoing charged hadrons (and neutral mesons) can be measured. $E_{\rm cal}$ is in general defined as
\begin{equation}
    E_{\rm cal} = E_{\ell'} + \sum T_{p} + \sum E_{\pi^{\pm}} + \sum E_{K^{\pm}} + \sum E_{\gamma} + \epsilon\, ,
\end{equation}
where $T_{p}$ is the kinetic energy of the measured outgoing protons, $E_{\pi^{\pm}} (E_{K^{\pm}})$ is the energies of the measured outgoing charged pions (kaons), the sum of photon energies, and $\epsilon$ is the average nucleon separation energy, taken here as 21 MeV.
In this particular e4$\nu$ analysis, since the signal is one proton and zero pions, $E_{\rm cal}$ is given by $E_{\rm cal} = E_{e'} + T_{p}  + \epsilon$.

We find in Fig.~\ref{fig:e4v1p0pi} that the peak near the beam energy is well described by \achilles and is dominated by quasi-elastic events. Resonance events contribute to smaller values of $E_{\rm cal}$ as they lead to knocked out neutrons or other undetected/under-threshold particles. Their contribution is diminished compared to the $0\pi$ selection because of the single proton requirement. Figure~\ref{fig:e4v1p0piPT} shows the comparison with the cross-section differential in the transverse momentum imbalance, $P_{T}$, defined by
\begin{equation}
    P_{T} = |\bm{p}_{T}^{e'} + \bm{p}_{T}^{p}|\, .
\end{equation}
$P_{T}$ belongs to a class of observables called \emph{Transverse Kinematic Imbalance} (TKI) variables which offer sensitivity to a range of nuclear affects. In contrast with the energy estimators, $P_{T}$ is more sensitive to the kinematics of the outgoing nucleons from $\pi N$ interactions in the INC.
Further information about $P_{T}$ and other TKI variables can be found in App.~\ref{sec:TKI}.
Figure~\ref{fig:e4v1p0piPT} shows small differences between the two cascade modes in the resonance contribution. We expect the additional strength needed to match the data will be provided by MEC and a small contribution from DIS.

\subsection{Comparison to T2K \label{sec:T2K}}

This section presents comparisons to measurements by the T2K collaboration for neutrino scattering on plastic scintillator $(C_8 H_8)$ in the T2K ND280 near detector with lan average beam energy of $\langle E_\nu\rangle \approx 600~{\rm MeV}$.
The measurements include both charged-current pionless (CC$0\pi$)~\cite{T2K:2018rnz} and charged-current single-pion (CC$1\pi^{+}$) events~\cite{T2K:2021naz}.
Events were generated with \achilles using the T2K beam flux from Ref.~\cite{T2K:2015sqm}. We note that the nucleon axial form factor used in these and all following comparisons is the $z$-expansion form factor fit to deuterium data~\cite{Meyer:2016oeg}.
A detailed study of uncertainties from the form factors is left for when \achilles includes all interaction channels~\cite{Simons:2022ltq}.
Comparisons were carried out by writing events to the \textsc{NuHepMC} file format~\cite{Gardiner:2023ejq} and processing them through the \textsc{nuisance} framework~\cite{Stowell:2016jfr}.

\begin{figure*}
    \centering
    \includegraphics[width=0.49\linewidth]{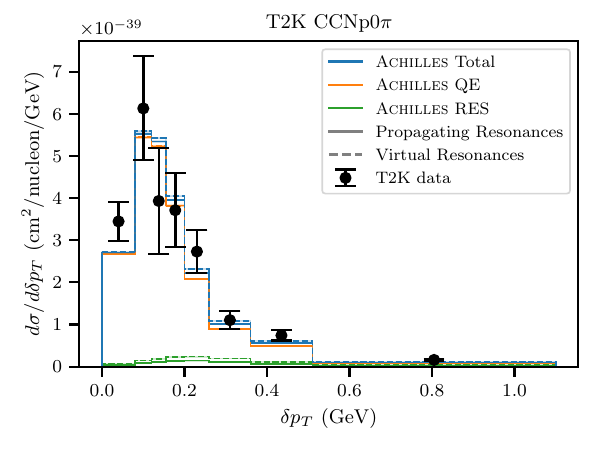}
    \includegraphics[width=0.49\linewidth]{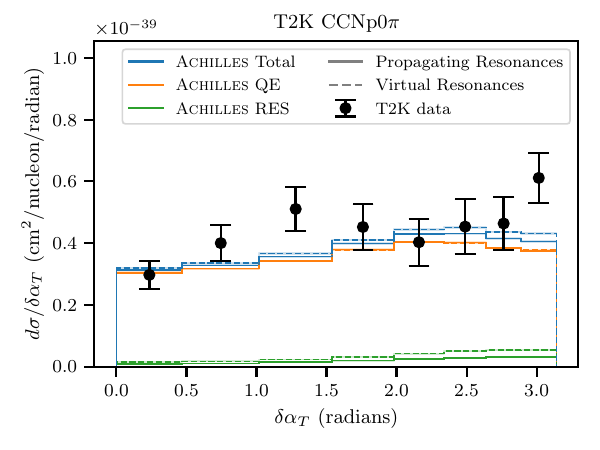}
    \caption{
    \textbf{Left:}
        Comparison of the CC$0\pi$ differential cross section with respect to the transverse momentum imbalance $\delta p_T$ measured by T2K~\cite{T2K:2018rnz} compared to predictions by \achilles.
    \textbf{Right:}
        Comparison of the CC$0\pi$ differential cross section with respect to the transverse boosting angle $\delta \alpha_T$ measured by T2K~\cite{T2K:2018rnz} compared to predictions by \achilles.
    }
    \label{fig:T2K_CC0pinp_STV_XSec_1Ddpt_alpha}
\end{figure*}

\Cref{fig:T2K_CC0pinp_STV_XSec_1Ddpt_alpha} shows comparisons to CCNp$0\pi$ cross-section measurements by T2K, differential in the TKI variables $\delta p_T$ and $\delta\alpha_T$ defined in App.~\ref{sec:TKI}. 
As expected for CC$0\pi$ events at the beam energies of T2K, the contribution from resonance scattering followed by pion absorption is small compared to the dominant quasi-elastic scattering.
The resonance contribution to the $0\pi$ sample in the Virtual Resonances cascade mode is larger than that from the Propagating Resonances mode, which is a reflection of the differences in the pion absorption cross section shown in~\cref{fig:c12_ar40_validation}.
The \achilles predictions are in agreement with the data, except at large $\delta p_{T}$ and large $\delta\alpha_{T}$. In these regions MEC producing a two-nucleon final state are expected to contribute at the $10\%$ level. Their inclusion will be the subject of future work and is expected to increase the agreement with the data. DIS interactions are expected to contribute at the percent level or below.

\begin{figure*}
    \centering
    \includegraphics[width=0.49\linewidth]{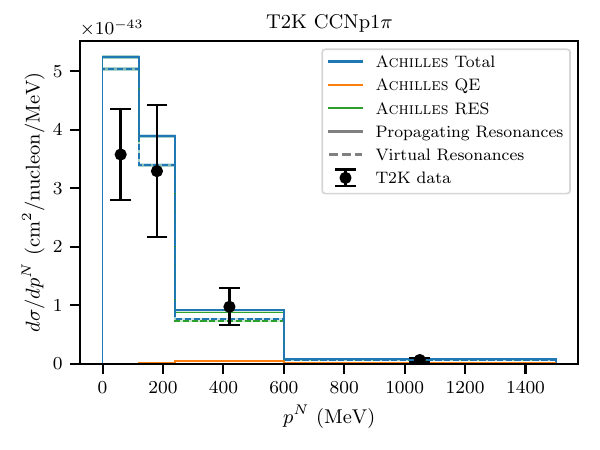}
    \includegraphics[width=0.49\linewidth]{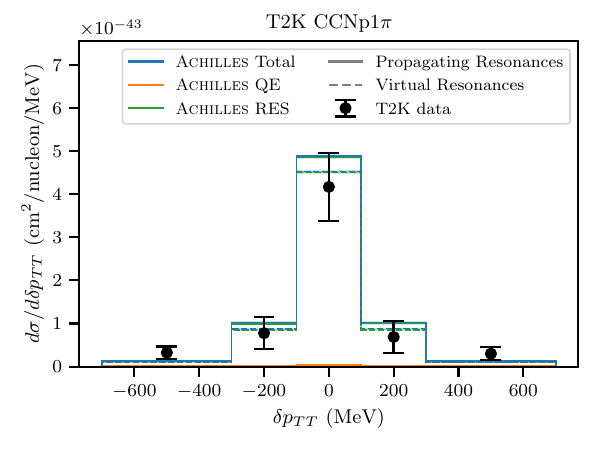}
    \caption{
    \textbf{Left:} Comparison of the CC$1\pi^{+}$ differential cross sections with respect to the initial nucleon momentum $p^N$ and \textbf{Right:} to the double transverse momentum imbalance $\delta p_{TT}$ measured by T2K~\cite{T2K:2021naz} 
    compared to predictions by \achilles.
    }
    \label{fig:T2K_CC1pipNp_CH_XSec_momenta}
\end{figure*}

\Cref{fig:T2K_CC1pipNp_CH_XSec_momenta} shows comparisons to CC$1\pi^{+}$ cross-section measurements by T2K differential in the TKI variables $p^N$ and $\delta p_{TT}$. 
The agreement with the data is excellent, with contributions from resonant scattering accounting for nearly the entire cross section. 
This data set is complementary, as pion absorption decreases the contribution of resonance events in this sample, whereas it leads to an increase of events in the $0\pi$ sample. 
In the $1\pi$ sample this is reflected in the larger cross section predicted by the Propagating Resonances vs. Virtual Resonances mode, again consistent with the different pion absorption cross sections predicted by these two modes. 
We note that the contribution from DIS is expected to be small at these energies and its inclusion will not dramatically alter our conclusions.

Additional comparisons to T2K data can be found in \cref{app:exp_compare}.

\subsection{Comparison to MINER$\nu$A \label{sec:MINERvA}}

This section presents comparisons to measurements by the MINER$\nu$A collaboration for neutrino scattering on plastic scintillator with an average beam energy of $\langle E_\nu \rangle \approx 3~{\rm GeV}$~\cite{MINERvA:2018hba}.
Ref.~\cite{MINERvA:2018hba} focused on CC$0\pi$ events with one or more final-state protons but zero final-state pions.
Events were generated with \achilles using the MINER$\nu$A beam flux from Ref.~\cite{MINERvA:2017okh}.
Comparisons were carried out by writing events to the \textsc{NuHepMC} file format~\cite{Gardiner:2023ejq} and processing them through the \textsc{nuisance} framework~\cite{Stowell:2016jfr}.

\begin{figure*}
    \centering
    \includegraphics[width=0.49\linewidth]{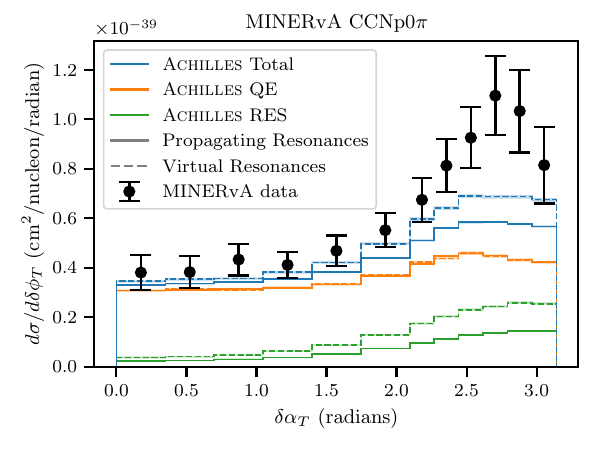}
    \includegraphics[width=0.49\linewidth]{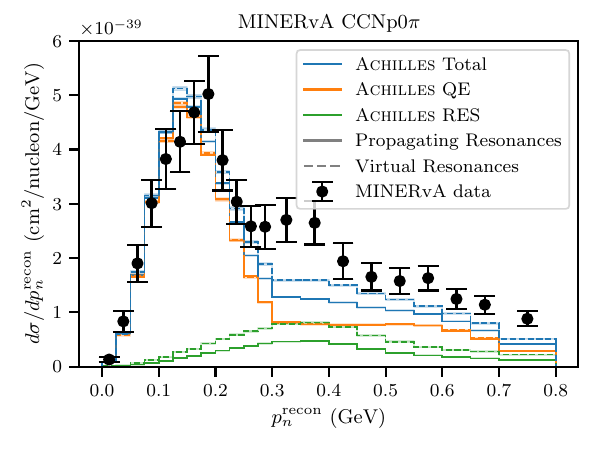}
    \caption{
    \textbf{Left:}
        Comparison of the CC$0\pi$ differential cross section with respect to the transverse boosting angle $\delta \alpha_T$ measured by MINER$\nu$A~\cite{MINERvA:2018hba} compared to predictions by \achilles.
    \textbf{Right:}
        Comparison of the CC$0\pi$ differential cross section with respect to the reconstructed initial neutron momentum $p_n^{\rm recon}$ measured by MINER$\nu$A~\cite{MINERvA:2018hba} compared to predictions by \achilles.
    }
    \label{fig:MINERvA_STV_alpha_pnrecon}
\end{figure*}

The neutrino beam at MINER$\nu$A's experimental setup in Ref.~\cite{MINERvA:2018hba} is at substantially higher energies than T2K in \cref{sec:T2K}.
\Cref{fig:MINERvA_STV_alpha_pnrecon} shows the cross sections for the transverse boosting angle $\delta\alpha_{T}$ and the reconstructed initial neutron momentum $p_{n}^{\rm{recon}}$ defined in App.~\ref{sec:TKI}. 
Although MINER$\nu$A's CC$0\pi$ measurement has a similar signal definition to T2K's, resonance scattering plays a much more important role at these energies. For example, at large $\delta\alpha_T$, roughly one third to a half of events come from resonance scattering. The absolute fractions for $p_{n}^{\rm{recon}} > 0.3$ GeV are similar. Interestingly at these energies the difference between the two cascade modes' predictions of the resonance contribution is enhanced, with the Virtual Resonances mode providing twice the predicted contribution from pion absorption when compared to the Propagating Resonances. This highlights the additional power of neutrino-nucleus scattering over hadron-nucleus scattering for constraining pion-nucleus interactions. 

In the region where quasi-elastic scattering is expected to dominate with only small contributions from FSI, i.e. at low $\delta\alpha_{T}$, the \achilles prediction is in excellent agreement with the data. The under prediction outside this region is due to the lack of MEC which is predicted to give a larger contribution at MINER$\nu$A beam energies than in T2K. MEC interactions shift the $\delta\alpha_{T}$ distribution towards higher values, due to the primary outgoing proton having only a fraction of the transferred momentum~\cite{Filali:2024vpy}. 
A similar discrepancy occurs at values of $\delta p_{T}$ and $p_{n}^{\rm{recon}}$ $>$ 0.3 GeV. Additional comparisons to MINER$\nu$A data can be found in \cref{app:exp_compare}.

We refrain from showing comparisons to MINER$\nu$A $N\pi$ data as the contributions from DIS are comparable to that from resonance scattering. 

\subsection{Comparison to MicroBooNE \label{sec:MicrBooNE}}
This final section presents comparisons to measurements by the MicroBooNE collaboration for neutrino scattering on liquid argon with an average beam energy of $\langle E_\nu \rangle \approx 0.8~{\rm GeV}$~\cite{MicroBooNE:2024pdj}, similar to T2K. Ignoring detector related effects, comparisons between T2K and MicroBooNE can be used to isolate $A$-dependent nuclear effects. We compare to two measurements from MicroBooNE: Ref.~\cite{MicroBooNE:2023tzj} which focused on CC1p$0\pi$ events with one final-state proton and zero final-state pions, and Ref.~\cite{MicroBooNE:2024pdj} which focused on NC$1\pi^{0}$ events with and without protons. 
Events were generated with \achilles using the MicroBooNE Booster Neutrino Beam (BNB) flux from Ref.~\cite{MicroBooNE:2024pdj}. 
All \achilles predictions have been folded with the corresponding smearing matrix, defined in Ref.~\cite{Tang:2017rob} as $A_{C}$, in order to bring our predictions into the regularized space used by MicroBooNE~\cite{MicroBooNE:2023tzj,MicroBooNE:2024pdj}.

\begin{figure*}[t!]
    \centering
    \includegraphics[width=\linewidth]{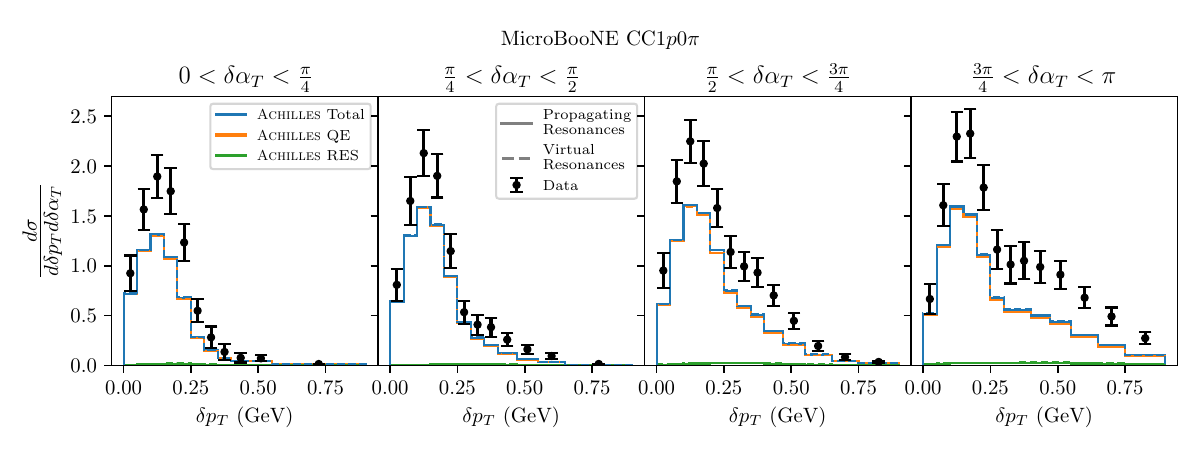}
    \caption{Comparison of CC1p$0\pi$ double-differential cross section measurement with respect to the transverse momentum $\delta p_T$ and the transverse boosting angle $\delta\alpha_T$ measured by MicroBooNE~\cite{MicroBooNE:2023tzj}
    in units of $10^{-39} {\rm cm}^{2}$/GeV/degree compared to predictions by \achilles.
    }
    \label{fig:1mu1p0pi}
\end{figure*}

\begin{figure*}[t]
    \centering
    \includegraphics[width=\linewidth]{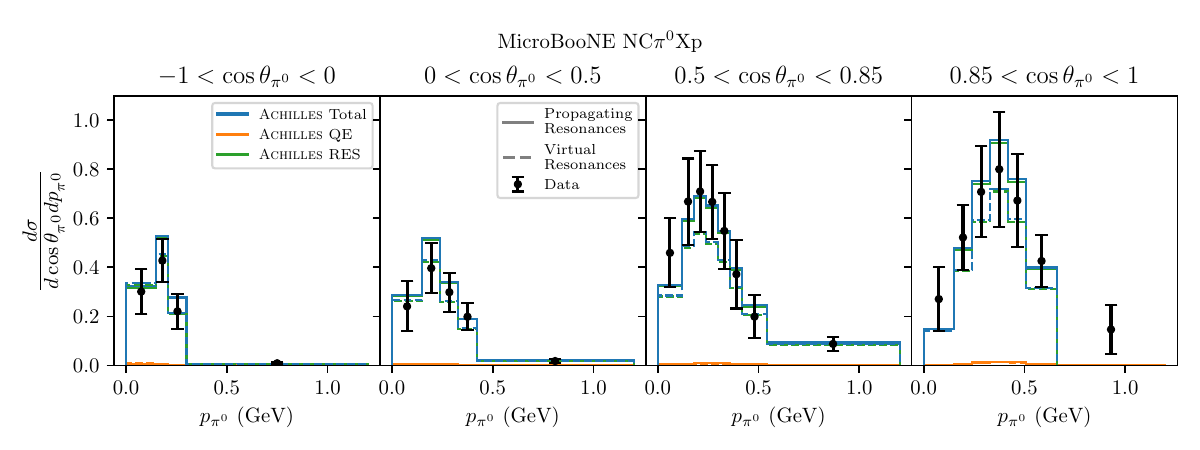}
    \caption{Comparison of NC1$\pi^{0}$ double-differential cross section measurement with respect to the scattering angle $\cos \theta_{\pi^0}$ and momentum $p_{\pi^{0}}$ momentum of the final-state $\pi^{0}$
    measured by MicroBooNE~\cite{MicroBooNE:2024sec} in units of $10^{-39} {\rm cm}^2$/GeV/nucleon compared to predictions by \achilles.
    }
    \label{fig:NCpi0Xpdoublediff}
\end{figure*}

Figure \ref{fig:1mu1p0pi} shows comparisons to  CC1p$0\pi$ cross-section measurements by MicroBooNE, double-differential in $\delta p_{T}$ and $\delta\alpha_{T}$.
Measurements of $\delta p_{T}$ where $\delta\alpha_{T}$ is small have been used to isolate nuclear structure effects, while at large $\delta\alpha_{T}$ FSI and inelastic interactions dominate.
\achilles reproduces the qualitative shape of the data in each bin, while the normalization is underpredicted.
This deficit is unsurprising for bins with moderate to large $\delta\alpha_{T}$ given the lack of MEC in the \achilles predictions.
The lack of strength in the lowest $\delta\alpha_{T}$ bin is notable, as this bin is supposed to have the smallest contribution from inelastic interactions.

Reference~\cite{Nikolakopoulos:2024mjj} studied the same type of events with several different INCs and hard-scattering models and found a similar under prediction.
We comment on several possible causes.
First is an increase in the axial form factor consistent with that found by Lattice QCD calculations~\cite{RQCD:2019jai,Park:2021ypf,Djukanovic:2022wru}, and a recent measurement by MINER$\nu$A~\cite{MINERvA:2023avz,Tomalak:2023pdi}.
Both point to an axial form factor which is over 20\% larger at $Q^{2} > 0.4$ $\rm{GeV}^{2}$ than that extracted from neutrino-deuterium scattering~\cite{Meyer:2016oeg}, leading to a 10-20\% increase in CC$0\pi$ cross sections at BNB energies.
However, this increase would have to be reconciled with high $Q^{2}$ data from other neutrino-nucleus cross section measurements. 
Second is a lack of interference between one- and two-body currents leading to one-nucleon knockout~\cite{Franco-Munoz:2022jcl,Lovato:2023khk,Franco-Munoz:2023zoa}.
In Ref.~\cite{Lovato:2023khk} this interference was shown to lead to a 10\% increase in the neutrino-nucleus cross section on carbon at MiniBooNE.
On an asymmetric nucleus like argon with more $np$ pairs than carbon, this enhancement could be even larger.
Lastly, we note that migration between bins caused by the $A_{C}$ matrix could mean that this low $\delta\alpha_{T}$ receives a large contribution from MEC when measured in the regularized space reported by MicroBooNE. This effect will be investigated further once MEC are included in \achilles predictions. 

\begin{figure}
    \centering
    \includegraphics[width=\linewidth]{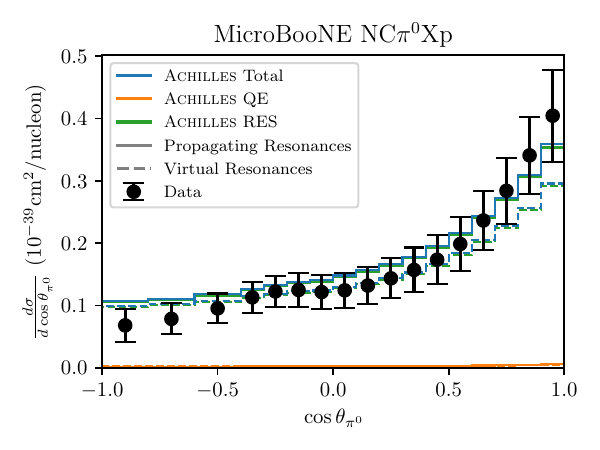}
    \caption{Comparison of NC1$\pi^{0}$ differential cross section measurement with respect to the $\pi^{0}$ scattering angle measured by MicroBooNE~\cite{MicroBooNE:2024sec} compared to predictions by \achilles.
    }
    \label{fig:NCpi0Xpcospi}
\end{figure}

Figures~\ref{fig:NCpi0Xpdoublediff} and~\ref{fig:NCpi0Xpcospi} show our last two comparisons to NC$\pi^{0}\rm{Xp}$ cross-sections from MicroBooNE, differential in $\pi^{0}$ scattering angle and double-differential in $\pi^{0}$ momentum. These events contain a single $\pi^{0}$ and any number of additional hadrons. Neutral current $\pi^{0}$ production is an important background in electron neutrino appearance measurements as the two photons from $\pi^{0}$ decay can mimic the electromagnetic shower from a single electron. We find that the \achilles prediction is in excellent agreement with the data for both the single and double differential cross sections. The Propagating Resonances cascade mode consistently predicts a cross-section 10 to 15\% larger than the Virtual Resonances mode. This is consistent with the differences between the two cascade modes' predictions for T2K CC$1\pi^{+}$ cross sections shown in Fig.~\ref{fig:T2K_CC0pinp_STV_XSec_1Ddpt_alpha}. This is most likely from the larger absorption cross section predicted by the Virtual Resonances mode, but could also stem from the different charge-exchange cross sections predicted by the DCC and GiBUU $\Delta$ models.

The exclusive predictions provided by the \achilles DCC implementation are key to obtaining the correct kinematics for the outgoing $\pi^{0}$. In the future, comparison of the cross-section split into 0p and Np final states will allow us to constrain the isospin dependence of the pion absorption models in \achilles. We note that \achilles does not currently include NC coherent pion production which contributes mostly at forward pion scattering angles, or DIS which is expected to be subdominant at BNB energies.

\section{Conclusions \label{sec:conclusions}}

The present work has extended the \achilles Monte Carlo event generator by incorporating single-pion production in a fully exclusive fashion.
The electroweak interaction vertex is modeled by combining the state-of-the-art Dynamical Coupled-Channels (DCC) approach with realistic hole spectral functions, which account for correlations in both the initial target state and the residual spectator system.
We have validated the hard interaction vertex by comparing to high precision inclusive electron scattering data from JLab, as well as bubble-chamber data from the ANL and BNL experiments that used elementary nucleon targets. 
These comparison serve to validate both the implementation of the DCC model in \achilles as well as its ability to describe quantitatively experimental data from both vector and axial-vector probes. 

The present work has also extended the \achilles INC to include pions using two different treatments of resonances to furnish an estimate of theoretical uncertainty.
In the first model, resonances have been fully integrated out so that the only propagating degrees of freedom are nucleons and pions; the required $\pi N$ scattering amplitudes are provided DCC and the Oset optical-potential models.
In the second model, all $\pi N$ interactions during the INC are mediated by propagating $\Delta$ resonances, which are produced and decayed as explicit propagating degrees of freedom along side pions and nucleons.
The two cascade modes qualitatively reproduce hadron-nucleus scattering observables like the total pion reaction cross section on carbon and the pion-absorption cross section on carbon and argon.
In some instances (e.g., the total reaction cross section of $\pi^+$ scattering on $^{12}$C for $p_\pi \gtrsim 0.5~{\rm GeV}$ or the $\pi^{+}-{}^{12}$C absorption cross section near the $\Delta$ peak), differences between the modes' predictions as well as differences with the data suggest that additional in-medium effects and interaction mechanisms in the cascade will be important to improve agreement with data.
Additional study of these effects will be the subject of future work.

We compared \achilles predictions to exclusive electron- and neutrino-nucleus scattering data for a range of beam energies and targets.
\Cref{sec:e4v} presented comparisons to recent measurements from CLAS and e4$\nu$ for electron-carbon scattering, focusing on quasielastic-like $1e1p0\pi$ events.
Similar comparisons were carried out previously in Ref.~\cite{Isaacson:2022cwh} including QE events only.
As expected, including single-pion production has improved agreement between \achilles' predictions and the experimental data.
\Cref{sec:T2K} presented comparisons to measurements of neutrino scattering on plastic scintillator at $\langle E_\nu \rangle \approx 0.6$ GeV by T2K including ${\rm CC}0\pi$ and ${\rm CC1\pi}$ events for TKI observables.
Few-sigma qualitative agreement in essentially all kinematic bins was observed.
\Cref{sec:MINERvA} also compared to neutrino scattering on plastic scintillator, this time for measurements at higher energies ($\langle E_\nu \rangle \approx 3$ GeV) from MINER$\nu$A.
As expected, resonance events make up a larger fraction of the total cross section, but there is also significant contributions from MEC and DIS which our prediction is currently missing.
At these energies, observables like the transverse boosting angle $\delta \alpha_T$ and reconstructed initial neutron momentum $p_n^{\rm recon}$ were found to be sensitive to difference in the pion mode (virtual resonances vs propagating resonances) used in the INC.
Finally, \cref{sec:MicrBooNE} gave comparisons to neutrino-argon scattering with $\langle E_\nu \rangle \approx 0.8$ GeV at MicroBooNE.
\achilles under-predicts the ${\rm CC}0\pi 1p$ double-differential cross section $d\sigma/d\delta p_{T}d\delta\alpha_T$, a feature which has been noted previously in predictions from other generators~\cite{Nikolakopoulos:2024mjj}.
Better agreement is observed for ${\rm NC}\pi^0{\rm X}p$ cross sections, where \achilles reproduces the experimental data at the $\approx 1\sigma$ level in essentially all kinematics bins.

Overall, the inclusion of single-pion production in \achilles represents an important milestone toward our goal of including all physical processes relevant for the accelerator-based neutrino program.
In the near future, we also plan to include support for MEC and DIS.
The results of the present work (e.g., in comparisons to MINER$\nu$A data) highlight the importance and utility of neutrino cross-section measurements to distinguish between different model treatments of nuclear effects.
It will be interesting and important to revisit these effects in the future once \achilles also includes MEC and DIS.

Finally, once all interaction mechanisms are present, providing a complete error estimate will be important to determine necessary paths for improvement to meet the needs of the DUNE experiment.

\begin{acknowledgments}
We would like to thank Afroditi Papadopoulou for sharing the details of the e4$\nu$ and MicroBooNE 1p$0\pi$ results and Ben Bogart for providing comparisons with the MicroBooNE NC$\pi^{0}$ measurement. 
We would like to thank Luke Pickering with help using the Nuisance analysis framework, and for help with the NuHepMC interface to Nuisance.
We would like to thank Afroditi Papadopoulou for comments on the manuscript.

This manuscript has been authored by Fermi Forward Discovery Group, LLC
under Contract No. 89243024CSC000002 with the U.S.\ Department of Energy,
Office of Science, Office of High Energy Physics.
The present research is supported by the U.S. Department of Energy, Office of Science, Office of Nuclear Physics, under contracts DE-AC02-06CH11357 (A.~L.), by the DOE Early Career Research Program (A.~L.), by the SciDAC-5 NeuCol program (J.~I., A.~L., and N.~R.), by the Neutrino Theory Network (A.~N.), and by the Nuclear Theory for New Physics Topical collaboration (A.~L.,  N.~R., and N.~S.).

This work was supported in part by Colorado State University through high-performance computer time and resources provided by the Data Science Research Institute.

Feynman diagrams were produced through the use of Tikz-Feynman~\cite{Ellis:2016jkw}.
Figures were generated using \textsc{matplotlib}~\cite{Hunter:2007} and \textsc{seaborn}~\cite{Waskom2021}.

The data in this paper can be reproduces using version 0.3.0 of Achilles~\cite{Isaacson2026-hg} with the run cards and Nuisance analyses found in Ref.~\cite{Isaacson2026-ly}.

\end{acknowledgments}

\onecolumngrid

\appendix

\section{Phase-space handling}
\label{app:phase_space}
The following is based closely on the TASI lecture notes from S. H\"oche~\cite{Hoche:2014rga}.
In general, the differential final-state phase space element for a $2 \to (n-2)$ scattering is
\begin{equation}
   {\rm d}\Phi_n\left(\{\vec{p}\}\right) = \left[ \prod_{i=3}^{n}\frac{{\rm d}^4p_i}{(2\pi)^3}\delta\left(p_i^2-m_i^2\right)\Theta\left(E_i\right)\right](2\pi)^4\delta^{(4)}\left(p_1+p_2-\sum_{i=3}^np_i\right)\,,
   \label{eq:ps_full}
\end{equation}
where $p_1$ and $p_2$ are initial state momenta, $m_i$ are the on-shell masses of the outgoing particles.
Generic techniques for handling high-dimensional phase-space integrals were proposed in Ref.~\cite{Byckling:1969sx}. The authors showed that it is possible to factorize the phase space into three components, which correspond to $2\to 2$ scattering, $1\to 2$ decay, and $2\to 1$ annihilation processes.
\cref{eq:ps_full} factorizes as~\cite{James:1968gu}
\begin{equation}
   {\rm d}\Phi_n(p_1,p_2;p_3,\ldots, p_n) = {\rm d}\Phi_{n-m+1}(p_1,p_2;p_3,\ldots,p_{n-m},P)\frac{{\rm d}P^2}{2\pi}{\rm d}\Phi_m(P;p_{n-m+1},\ldots,p_n)\,,
   \label{eq:ps_factorize}
\end{equation}
where $P$ denotes a virtual intermediate particle. While the particle $P$ has no direct physical interpretation, it may be associated with an $s$-channel propagator formed by the set of external states $\{p_{n-m+1},\ldots,p_n\}$.
Leveraging knowledge of the Feynman diagrams involved in the matrix element allows one to efficiently map out the peak structure of certain diagrams squared. Therefore, the ability to match $P$ to a propagator in the Feynman diagram enables efficient Monte-Carlo sampling of the integral. However, each diagram can have many different combinations of propagator structures, leading to the need to use the multi-channel method to find the optimal integrator over the full matrix element squared~\cite{Kleiss:1994qy}.
Repeating the process in \cref{eq:ps_factorize} allows a decomposition of the complete phase space into three elementary building blocks defined as
\begin{align}
    {\rm d}\Phi_2(p_a,p_b;p_i,p_j) = \frac{\lambda(s_{ab},s_i,s_j)}{16\pi^2 2s_{ab}} {\rm d}\cos\theta_i{\rm d}\phi_i\,,\nonumber \\
    {\rm d}\Phi_2(p_ij;p_i,p_j) = \frac{\lambda(s_{ij},s_i,s_j)}{16\pi^2 2s_{ij}} {\rm d}\cos\theta_i{\rm d}\phi_i\,, \\
    {\rm d}\Phi_1(p_a,p_b;p_i) = (2\pi)^4{\rm d}^4p_i \delta^{(4)}(p_a+p_b-p_i)\,,\nonumber
    \label{eq:ps_building_blocks}
\end{align}
where $\lambda$ is the K\"all\'en function.
The different building blocks can be associated with $t-$ and $s-$channel vertices, while the ${\rm d}P^2/2\pi$ can be associated with a propagator.
This enables the usage of techniques developed for automatic matrix element calculations to also be used for generating the phase space, as is the case in the automated matrix element calculator \textsc{Comix}~\cite{Gleisberg:2008fv}.

\section{Meson-baryon scattering amplitudes \label{app:MB_amplitudes}}

We express the cross section for meson-baryon scattering
\begin{equation}
m_i(\mathbf{k}_i) + B_i(-\mathbf{k}_i) \rightarrow m_f(\mathbf{k}_f) + B_f(-\mathbf{k}_f)
\end{equation}
with $\mathbf{k}_i,~\mathbf{k}_f$ the momenta in the center-of-mass (CM) frame, in terms of standard partial-wave amplitudes (PWA) $\tau^{L,\pm,I}_{i,f}$.
The PWA have well-defined pion-nucleon orbital angular momentum $L$, total angular momentum $J = L \pm 1/2 $, and total isospin $I$. The indices $i,f$ denote the initial, final meson baryon pair $\pi N, \eta N, K\Lambda, K \Sigma$.

From isospin conservation the PWA for each channel $i,f$
can be decomposed in terms of total isospin $I$ of the meson-baryon pair.
I.e $I = 1/2$ or $3/2$ for $\pi N$ and $K \Lambda$, and $I = 1/2$ for $\eta N$ and $K \Sigma$.
The PWA for a specific process $m B \rightarrow m^\prime B^\prime$, are then given by
\begin{equation}
    \tau^{L,\pm}_{mB,m^\prime B^\prime} = \sum_I \left( I^{m}, I^{m}_3 ; I^{B}, I^{B}_3 \vert I, I_3 \right)\left( I, I_3 \vert I^{m^\prime}, I^{m^\prime}_3 ; I^{B^\prime}, I^{B^\prime}_3 \right) \tau_{i, f}^{L,\pm,I}
\end{equation}
where $I^{m/B}, I_3^{m/B}$ denote the total isospin and its projection of the meson/baryon. The Clebsch-Gordan coefficients $\left( I^{m}, I^{m}_3 ; I^{B}, I^{B}_3 \vert I, I_3 \right)$ couple the isospin of the meson-baryon pair in the initial and final state to total isospin $I$.
In the following we express the cross section for a specific process in terms of PWA. We drop the subscript $mB, m^\prime B^\prime$ for ease of notation.

The differential cross section in the CM frame can be written in terms of two amplitudes
\begin{equation}
\label{eq:CSddif}
\frac{\mathrm{d} \sigma}{\mathrm{d} \Omega} = \frac{1}{\rvert \mathbf{k}_i\lvert^2} \rho_i \rho_f \left( \lvert T_{++} (s, \Omega) \rvert^2 + \lvert T_{+-}(s,\Omega) \rvert^2 \right) = \frac{\lvert \mathbf{k}_f \rvert}{\lvert \mathbf{k}_i \rvert} \left( \lvert h(s,\Omega) \rvert^2 + \lvert g(s,\Omega) \rvert^2 \right). 
\end{equation}
Here $T_{++}, T_{+-}$ are helicity flip and non-flip amplitudes, $h,g$ are spin-flip and non-flip amplitudes respectively. 
The phase-space factors for initial and final state $\rho_i, \rho_f$ are given by
\begin{equation}
    \rho(k) = \pi\frac{k}{W} E_m E_B
\end{equation}
with $E_m, E_B$ the CM energy of the meson and baryon, respectively, and $k$ the CM momentum. 
The helicity amplitudes are expanded in terms of the total angular momentum of the meson-baryon system $J$
\begin{align}
\label{eq:angular_dep}
T_{\lambda^\prime, \lambda}(s, \Omega) =& \sum_{J} (J+1/2)  T^{J}_{\lambda,\lambda^\prime}(s)e^{i(\lambda - \lambda^\prime)\phi}d_{\lambda, \lambda^\prime}^{J}(\theta).
\end{align}
All angular dependence of \cref{eq:CSddif} is given by the Wigner small-$d$ functions. 
For helicities $\lambda = \pm 1/2$ they are related to Legendre polynomials $P_l$ as
\begin{equation}
    (J+1/2)d^{J}_{\pm\frac{1}{2},\frac{1}{2}}(\theta) = \sqrt{\frac{1\pm\cos\theta}{2}}\left( P^\prime_{J+1/2}(\cos\theta) \mp P^{\prime}_{J-1/2}(\cos\theta) \right),
\end{equation}
where the primes denote the derivative with respect to $\cos\theta$.
The relation between the helicity amplitudes and amplitudes with definite parity, and hence $L$, is
\begin{equation}
    2T^{L,\pm}(s) = T_{++}^{J= L \pm 1/2}(s) \pm T_{+-}^{J = L \pm 1/2}(s).
\end{equation}
The (dimensionless) PWA are then defined as 
\begin{equation}\label{eq:dimensionless-PWAs}
    \tau^{L,\pm}(s) = -\sqrt{\rho_i \rho_f}~T^{L,\pm}(s).
\end{equation}
We have explicitly
\begin{equation}
    \label{eq:Tpp_multipoles}
    \sqrt{\rho_i\rho_f}~T_{+\pm}(s,\theta ) = \sum_J \left( J + 1/2 \right) d^{J}_{\pm\frac{1}{2},\frac{1}{2}}(\theta) \left( \tau^{J-1/2, +}(s) \pm \tau^{J+1/2, -}(s) \right),
\end{equation}
The differential cross section of Eq.~(\ref{eq:CSddif}) is then most conveniently expressed as
\begin{equation}
\label{eq:sigma_dif2}
       \frac{\mathrm{d}\sigma}{\mathrm{d}\Omega} = \frac{1}{\lvert \mathbf{k}_i \rvert^2} \Big\lvert \sum_L P_L(\cos\theta) \left[ L~\tau^{L, -}(s) + (L+1)~\tau^{L,+}(s)\right] \Big\rvert^2 + 
 \frac{\sin^2\theta}{\lvert\mathbf{k}_i\rvert^2} \Big\lvert \sum_L~ P_L^\prime(\cos\theta) \left[\tau^{L,+}(s) - \tau^{L,-}(s)\right] \Big\rvert^2.
\end{equation}
Here the first and second terms are proportional to the spin-flip and non-flip amplitudes $\lvert h \rvert^2, \lvert g\rvert^2$ respectively.

Efficient sampling of the CM scattering angle is readily obtained via the inverse sampling transform using the cumulative distribution function.
To do so, we express the probability density function for the scattering angle, given by the normalized angular distribution of Eq.~(\ref{eq:sigma_dif2}), in a monomial basis. 
This is done using the expansion coefficients $G_{k}^{l,n}$ and $H_k^{l,n}$ defined as
\begin{align}
\sum_{k = 0}^{l+n} H_k^{l,n} x^k &\equiv (1-x^2) P_l^{\prime}(x) P_n^\prime(x),\\
\sum_{k=0}^{l+n} G_k^{l,n} x^k &\equiv P_l(x) P_n(x).
\end{align}
These coefficients are multiplied by the PWA as in Eq.~(\ref{eq:sigma_dif2}), yielding the angular distribution. The cumulative distribution function is the integral of the resulting polynomial, given analytically in terms of the same coefficients.

From the orthogonality of the associated Legendre polynomials the angle-integrated cross section is given by squared PWA
\begin{equation}
\label{eq:sigma_int}
    \sigma(W) = \frac{4\pi}{\lvert \mathbf{k}_i \rvert ^2} \sum_L \left(~ L \lvert \tau^{L,-}(s) \rvert^2 + (L+1) \lvert \tau^{L,+}(s) \rvert^2 \right).
\end{equation}

\begin{figure}[t!]
    \centering
    \includegraphics[width=0.49\linewidth]{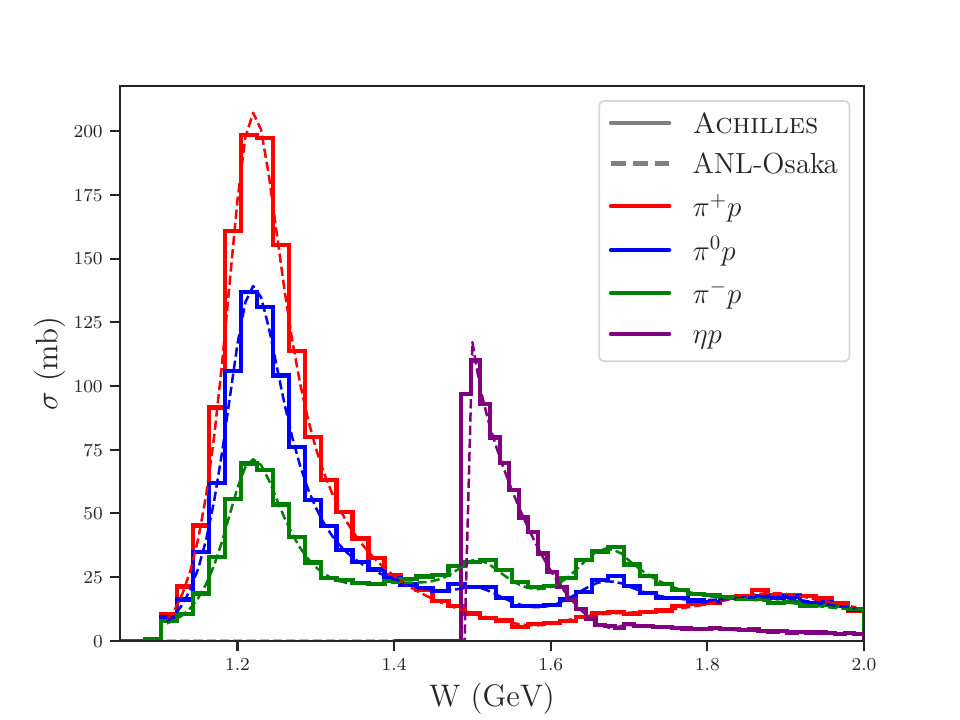}
    \includegraphics[width=0.49\linewidth]{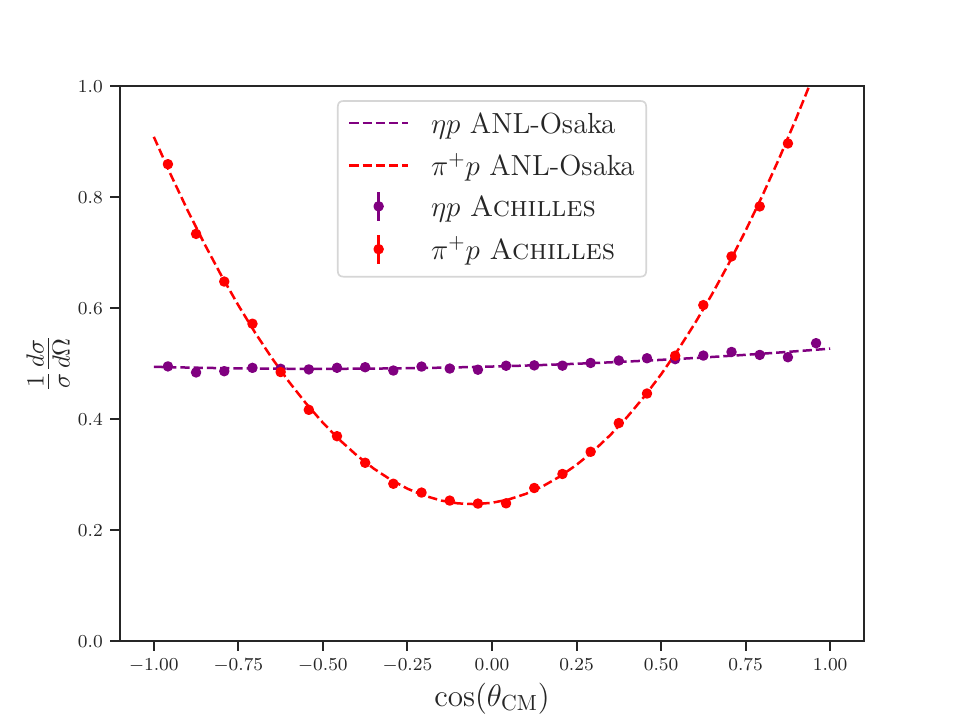}
    \caption{
    Validation of the \textsc{Achilles} implementation of the ANL-Osaka DCC Model.
    \textbf{Left:} Angle integrated cross sections for $\pi^+$ (red), $\pi^0$ (blue), $\pi^-$ (green), and $\eta$ (purple) scattering off the proton. Results only include the contributions for meson-baryon final-state channels $\pi N, \eta N, K \Lambda, K \Sigma$. Results shown by histograms are computed from the \achilles INC, while the dashed lines show the calculation directly from \cref{eq:sigma_int}.
    \textbf{Right:} Angular distributions in the CM frame for $\pi^+$ (red), and $\eta$ (purple) scattering off the proton at $p_{\pi/ \eta} = 300$ MeV. Results shown by points are computed from the \achilles INC, while the dashed lines show the calculation from the ANL-Osaka model. The angular distributions for $\pi^0$ and $\pi^{-}$ are identical to $\pi^{+}$ at these kinematics.
    }
    \label{fig:achilles-dcc-comparison}
\end{figure}

\section{Validation of Meson-Baryon Scattering}
\label{app:MBvalidation}
We validate the results for both the total cross sections, and angular distributions from the ANL-Osaka DCC model for meson-proton scattering by running the INC simulation for a stationary proton target.
Results are shown in \cref{fig:achilles-dcc-comparison} for incoming pions and eta mesons, validating the agreement between the \achilles INC and the original ANL-Osaka model. 
For the angular distributions in on the right of \cref{fig:achilles-dcc-comparison}, the pion isospin channels not shown have almost exactly the same angular distribution at this momenta, as this kinematic point is almost exactly on top of the $\Delta$ peak, where the different pion cross sections differ only by isospin factors. 
Results for neutron and Hyperon targets can be obtained in a similar manner.

\begin{figure}[t!]
    \centering
    \includegraphics[width=0.52\linewidth]{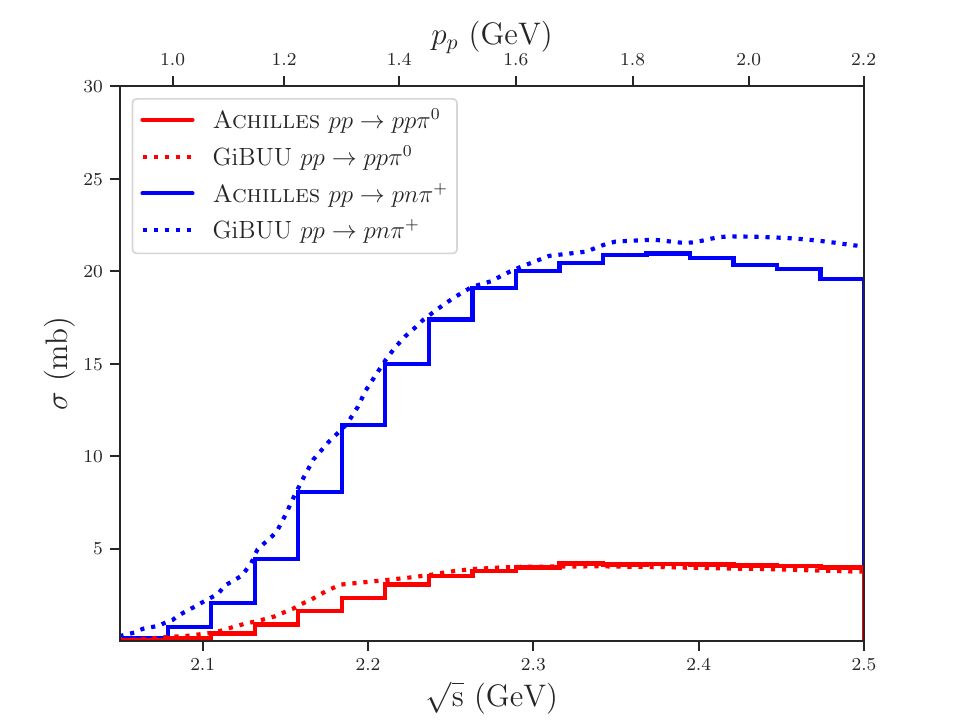}
    \includegraphics[width=0.47\linewidth]{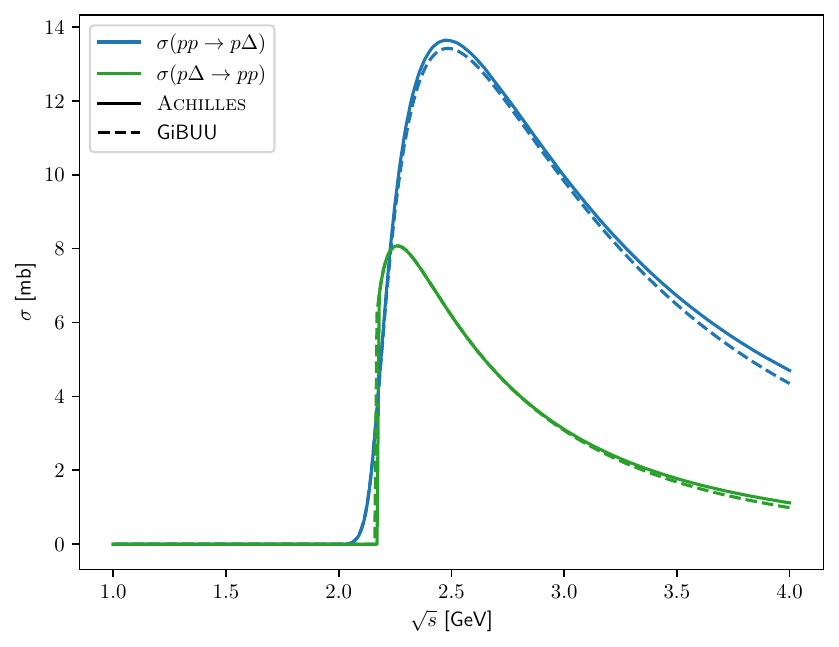}
    \caption{
    Comparison of results between \textsc{Achilles} and GiBUU.
    \textbf{Left:}
        Pion production cross sections versus $\sqrt{s}$ and $p_{beam}$ for $pp\rightarrow pn\pi^{+}$ (blue) and $pp\rightarrow pp\pi^{0}$ (red) in \achilles compared to the parameterized results from GiBUU~\cite{Buss:2011mx}.
    \textbf{Right:}
        Comparison of $\Delta$ cross sections described in Eqs.~\eqref{eq:detail_balance},~\eqref{eq:nntondelta} against the cross sections obtained from GiBUU~\cite{Buss:2011mx}.
    }
    \label{fig:pion_prod_validation}
\end{figure}

As a final check, Fig.~\ref{fig:pion_prod_validation} shows the total cross sections for charged and neutral pion production off of a hydrogen target using the results from the GiBUU $\Delta$ model. To compare with our predictions, we utilize the parameterizations of the total cross sections from GiBUU~\cite{Buss:2011mx}, which have been fit to global $NN\rightarrow NN\pi$ scattering data. Here one can see that the predictions from \achilles lie slightly below the data as we only include contributions from the resonant process $NN\rightarrow N\Delta\rightarrow NN\pi$. The missing strength is given by the non-resonant, background $NN\rightarrow NN\pi$ contributions which are for now not included in \achilles. 

\section{Delta Matrix Elements}
\label{app:delta_mat}

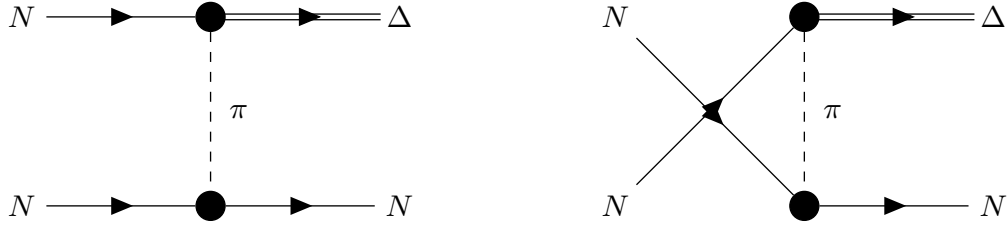
\begin{figure}
    \scalebox{1.25}{\begin{tikzpicture}
        \begin{feynman}
            \vertex (a) at (-2, 1) {$N$};
            \vertex[draw,circle,minimum size=0.2cm,fill] (b) at (0, 1) {};
            \vertex (b1) at (0, 0.97);
            \vertex (b2) at (0, 1.03);
            \vertex (c) at (1.8, 0.97);
            \vertex (c2) at (1.8, 1.03);
            \vertex (d) at (-2, -1) {$N$};
            \vertex[draw, circle, minimum size=0.2cm,fill] (e) at (0, -1) {};
            \vertex (f) at (2, -1) {$N$};
            \node[isosceles triangle, draw, fill, inner sep=0, minimum size=1.75mm] (T) at (1, 1) {}; 
            \node at ($(e)!0.5!(b)+(0.3,0)$) {$\pi$};
            \node at ($(f)+(0,2)$) {$\Delta$};

            \diagram* {
                (a) -- [fermion] (b),
                (b1) -- (c),
                (b2) -- (c2),
                (b) -- [scalar] (e),
                (d) -- [fermion] (e) -- [fermion] (f),
            };
        \end{feynman}
    \end{tikzpicture}}
    \hspace{2cm}
    \scalebox{1.25}{\begin{tikzpicture}
        \begin{feynman}
            \vertex (a) at (-2, 1) {$N$};
            \vertex[draw,circle,minimum size=0.2cm,fill] (b) at (0, 1) {};
            \vertex (b1) at (0, 0.97);
            \vertex (b2) at (0, 1.03);
            \vertex (c) at (1.8, 0.97);
            \vertex (c2) at (1.8, 1.03);
            \vertex (d) at (-2, -1) {$N$};
            \vertex[draw, circle, minimum size=0.2cm,fill] (e) at (0, -1) {};
            \vertex (f) at (2, -1) {$N$};
            \node[isosceles triangle, draw, fill, inner sep=0, minimum size=1.75mm] (T) at (1, 1) {}; 
            \node at ($(e)!0.5!(b)+(0.3,0)$) {$\pi$};
            \node at ($(f)+(0,2)$) {$\Delta$};

            \diagram* {
                (a) -- [fermion] (e),
                (b1) -- (c),
                (b2) -- (c2),
                (b) -- [scalar] (e),
                (d) -- [fermion] (b),
                (e) -- [fermion] (f),
            };
        \end{feynman}
    \end{tikzpicture}}
    \caption{Tree-level $\Delta$ production diagrams arising from the Lagrangian \cref{eq:DeltaProductionLagrangian}.
    A solid lines denotes a nucleon, the double lines denotes a $\Delta$, and a dotted line denotes a pion.
    }
    \label{fig:delta_diagrams}
\end{figure}

\Cref{fig:delta_diagrams} shows the diagrams in the tree-level calculation for $pp\to n \Delta^{++}$; other channels follow using isospin symmetry and Clebsch--Gordan coefficients.
The cross section is sum of three contributions: a direct term, a crossed term, and their interference.
The direct contribution is given by
\begin{align}
    \lvert\mathcal{M}({\rm direct})\rvert^2=z(t,\mu_\Delta)\left[\frac{f_\pi f_\pi^\ast F^2(t)}{m_\pi^2(t-m_\pi^2)}\right]^2 \frac{4m_N^2 t}{3\mu_\Delta^2} 
    \times \left[t - \left(\mu_\Delta-m_N\right)^2\right]\left[\left(\mu_\Delta+m_N\right)^2-t\right]^2 \,.
\end{align}
The crossed contribution follows from the exchange of the Mandelstam variables $t\leftrightarrow u$.
Finally, the interference term is given as
\begin{align}
\begin{split}
    2\textrm{Re}[\mathcal{M}({\rm direct}) &\mathcal{M}^\ast({\rm crossed})] =
    \left[
        \frac{f_\pi f_\pi^\ast F^2(t)}{m_\pi^2(t-m_\pi)^2}
    \right]
    \left[
        \frac{f_\pi f_\pi^\ast F^2(u)}{m_\pi^2(u-m_\pi)^2}
    \right]
    \sqrt{z(t,\mu_\Delta)z(u,\mu_\Delta)}
    \left(
        \frac{2m_N^2}{\mu_\Delta^2}
    \right)\\
    &\times \Bigg\{
    \left[
        tu+\left(\mu_\Delta^2-m_N^2\right)\left(t+u\right)-\mu_\Delta^4+m_N^4
    \right]
    \left[       
        tu+m_N\left(\mu_\Delta+m_N\right)\left(\mu_\Delta^2-m_N^2\right)
    \right]\\
     &-\frac{1}{3}\left[
        tu-\left(\mu_\Delta+m_N\right)^2\left(t+u\right)+\left(\mu_\Delta+m_N\right)^4
    \right]
    \left[
        tu-m_N\left(\mu_\Delta-m_N\right)\left(\mu_\Delta^2-m_N^2\right)
    \right]\Bigg\}\,.
\end{split}
\end{align}

The comparison between the implementation and the calculation within GiBUU can be seen in the right panel of Fig.~\ref{fig:pion_prod_validation}. The agreement between the two implementations are at an acceptable level and may differ from exact values of couplings and masses.

The matrix element for the process $N\Delta\to N\Delta$ is
\begin{align}
\begin{split}
    |\mathcal{M}_{N\Delta_i\to N\Delta_f}|^2&=\mathcal{I} \frac{1}{8}\left(\frac{f_{NN\pi}f_{\Delta\Delta\pi}}{m_\pi}\right)^2\frac{F^4(t)}{(t-m_\pi^2)^2}
    \times \frac{16(\mu_i + \mu_f)^2+m_N^2 t}{9\mu_i^2\mu_f^2} 
    \times \left(-\mu_i^2+2\mu_i\mu_f-\mu_f^2+t\right)\\
    &\times \left(\mu_i^4-2\mu_i^3\mu_f+12\mu_i^2\mu_f^2-2\mu_i\mu_f^3+\mu_f^4\right.
    \left. -2\mu_i t+2\mu_i\mu_f t -2\mu_f^2 t + t^2\right) .
\end{split}
\end{align}

In the above equation, $F(t)$ is the same form factor defined in \cref{eq:delta_formfactor}, $\mu_i$ and $\mu_f$ are the masses of the off-shell $\Delta$ resonance in the initial and final state, respectively.
The $NN\pi$ coupling constant $f_{NN\pi}=0.946$ and the $\Delta\Delta\pi$ coupling constant is given as $f_{\Delta\Delta\pi} = 9/5 f_{NN\pi}$. Finally, the isospin factors ($\mathcal{I}$) are given in Tab.~\ref{tab:delta_isospin}.

\begin{table}[t!]
    \centering
    \caption{Isospin factors for the process $N\Delta\to N\Delta$. The neutron channels are given by isospin symmetry.
    \label{tab:delta_isospin}
    }
    \begin{tabular}{c|c}
    \hline\hline
     Process    & Isospin Factor ($\mathcal{I}$)  \\
    \hline
    $p\Delta^{++} \to p\Delta^{++}$     & 9/4 \\
    $p\Delta^{+} \to n\Delta^{++}$     & 3 \\
    $p\Delta^{+} \to p\Delta^{+}$     & 1/4 \\
    $p\Delta^{0} \to p\Delta^{0}$     & 1/4 \\
    $p\Delta^{0} \to n\Delta^{+}$     & 4 \\
    $p\Delta^{-} \to p\Delta^{-}$     & 9/4 \\
    $p\Delta^{-} \to n\Delta^{0}$     & 3 \\
    \hline\hline
    \end{tabular}
\end{table}

\section{Transverse Kinematic Imbalance}\label{sec:TKI}
Transverse kinematic imbalance (TKI) variables are are popular observables designed for sensitivity to nuclear effects in neutrino-nucleus scattering~\cite{Lu:2015tcr}.
To establish notation, consider CCNp$0\pi$ scattering events $\nu+A\to \ell+p + X$, where $\ell$ and $p$ are respectively the outgoing lepton and leading proton.
The transverse momentum $\delta \bm{p}_T$ is defined via the vector sum of the outgoing transverse lepton and leading proton momentum
\begin{align}
    \delta\bm{p}_T
    &\equiv \bm{p}_T^{\ell'} + \bm{p}_T^{\rm N'}\,.
\end{align}
The notation $\delta\bm{p}_T$ is meant as a visual reminder that this transverse momentum would vanish in the absence of nuclear effects. 
As for any three-vector, $\delta\bm{p}_T$ is completely specified by its magnitude and two angles,
\begin{align}
    \delta p_T &\equiv | \bm{p}_T^{\ell'} + \bm{p}_T^{N'}|\\
    \delta \alpha_T &\equiv \arccos \left( \frac{-\bm{p}_T^{\ell'}\cdot\delta\bm{p}_T}{p_T^{\ell'} \, \delta p_T}\right)\\
    \delta\phi_T &\equiv \arccos \left( \frac{-\bm{p}_T^{\ell'}\cdot \bm{p}_T^{N'}}{p_T^{\ell'}  p_T^{N'}}\right),
\end{align}
where $p_T^{\ell'}$ and $p_T^{N'}$ are the projections of the momentum of the \emph{outgoing} lepton and leading proton onto the transverse plane defined by the incoming neutrino momentum.
The angle $\delta \alpha_T$ is referred to as the transverse boosting angle, while the angle $\delta \phi_T$ is the transverse deflection angle.
Ref.~\cite{Lu:2015tcr} discusses physical expectations for these observables and provides a diagram of the kinematic setup. Note that these same variables are used in electron-nucleus scattering as well, often denoted as $\bm{p}_T$.

Doubly transverse variables~\cite{Lu:2015hea,Lu:2015vri} can be defined analogously for CCNp$1\pi^{+}$ events
$\nu+A\to \ell'+p'+\pi' + X$.
A double-transverse axis $\bm{z}_{TT}$ is defined with respect to the plane of the initial- and final-state lepton momenta,
\begin{align}
    \hat{\bm{z}}_{TT} \equiv \frac{\bm{p}^\nu \times \bm{p}^{\ell'}}{|\bm{p}^\nu \times \bm{p}^{\ell'}|}.
\end{align}
Projecting the resonance momentum $\bm{p}^{N'}+\bm{p}^{\pi'}$, with $\bm{p}^{N'}$ the leading proton, onto the double-transverse axis gives the double-transverse momentum imbalance,
\begin{align}
    \delta p_{TT} &\equiv \left(\bm{p}^{N'}+\bm{p}^{\pi'}\right) \cdot \hat{\bm{z}}_{TT}\,.
\end{align} 
The magnitude of the initial nucleon momentum $p^N$ is defined in terms of it's components transverse and longitudinal to the initial neutrino momentum,
\begin{align}
    p^N = \sqrt{\left(\delta p_T^{\rm{CC}1\pi}\right)^2 + \left(p_L^{\rm{CC}1\pi}\right)^2}\,.
\end{align}
The transverse momentum in CCNp$1\pi^{+}$ events is given by
\begin{align}
    \delta \bm{p}_T^{\rm{CC}1\pi} = \bm{p}_T^{\ell'} + \bm{p}_T^{N'} + \bm{p}_T^{\pi'}\,.
\end{align}
The longitudinal component $p_L$ can similarly be computed in terms of final-state observables, although the expression is somewhat longer and will not be needed here. 
Details can be found in Ref.~\cite{T2K:2021naz}.

\section{Additional Experimental Comparisons} 
\label{app:exp_compare}

\begin{figure}[h!]
    \centering
    \includegraphics[width=0.45\linewidth]{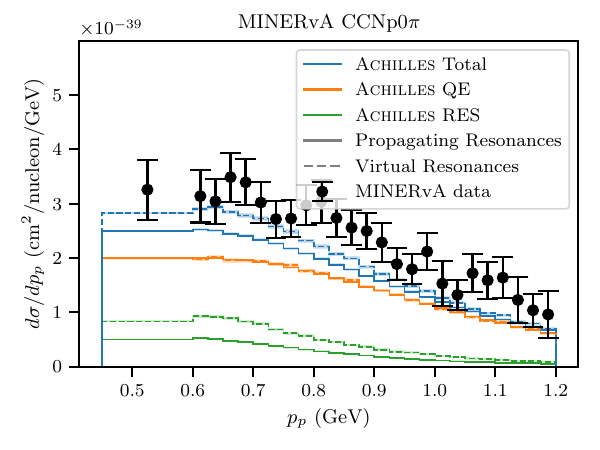}
    \includegraphics[width=0.45\linewidth]{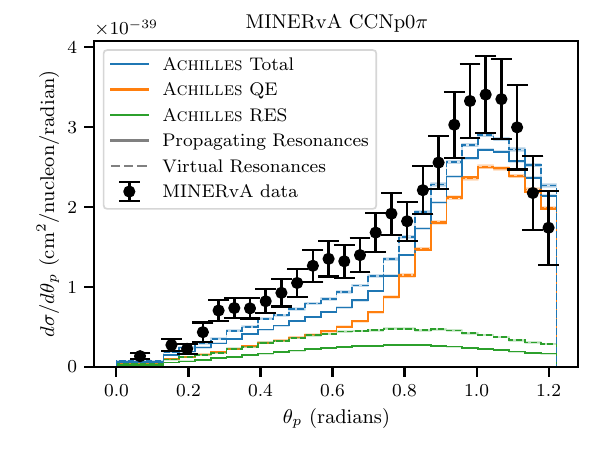}
    \includegraphics[width=0.45\linewidth]{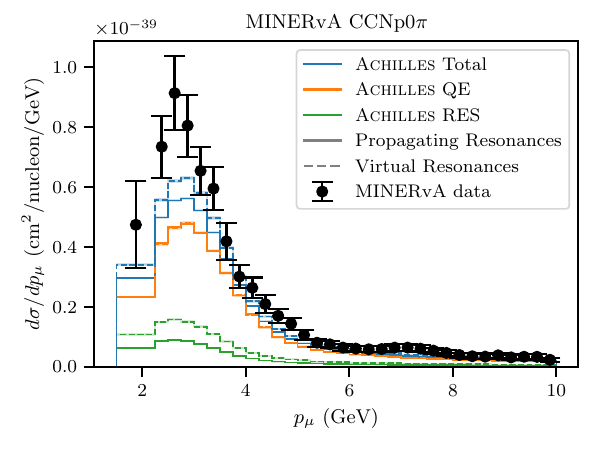}
    \includegraphics[width=0.45\linewidth]{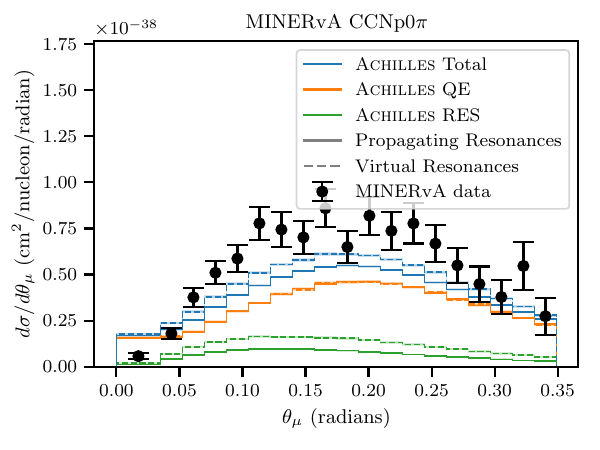}
    \includegraphics[width=0.45\linewidth]{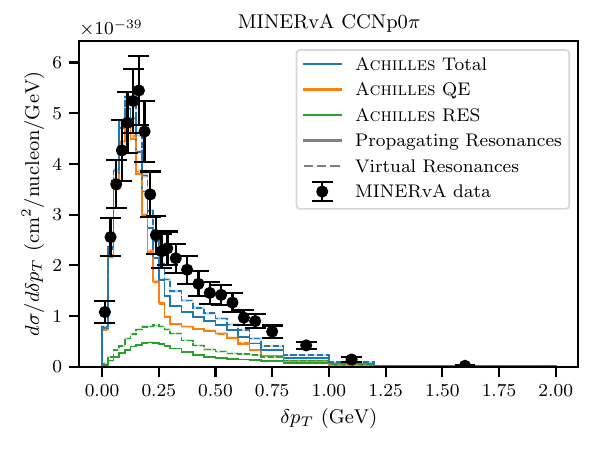}
    \includegraphics[width=0.45\linewidth]{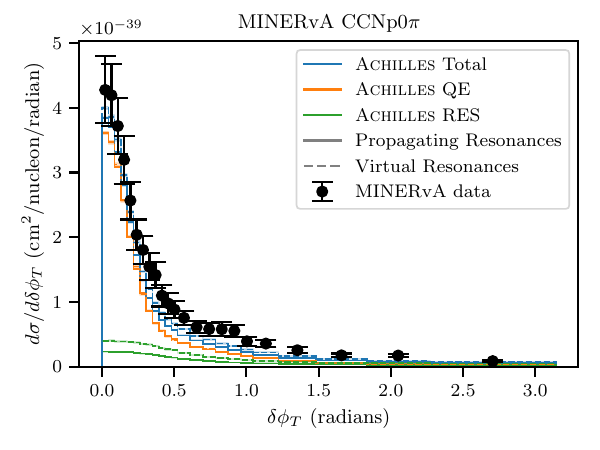}
    \caption{
    Comparison of CC$0\pi$ differential cross sections measured by MINER$\nu$A~\cite{MINERvA:2018hba} compared to predictions by \achilles.
    In the first row, the cross sections are differential in momentum $p_p$ and scattering angle $\theta_p$.
    In the second row, the cross sections are differential in 
    momentum $p_\mu$ and scattering angle $\theta_\mu$ of the outgoing muon.
    In the third row, the cross sections are differential in 
    the transverse momentum imbalance $\delta p_T$ and the transverse deflecting angle $\delta\phi_T$.
    } 
    \label{fig:MINERvA_extras}
\end{figure}

\begin{figure}
    \centering
    \includegraphics[width=0.49\linewidth]{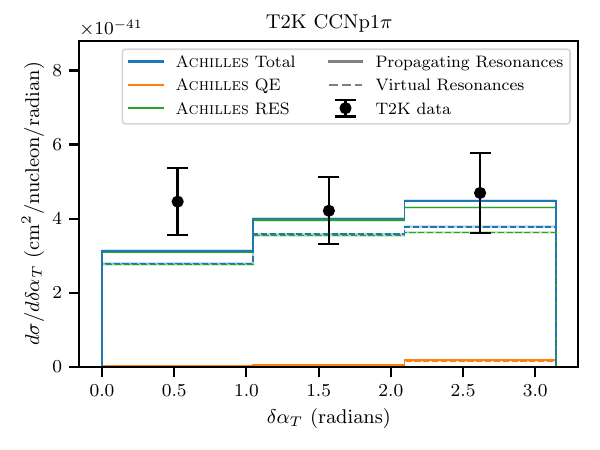}
    \includegraphics[width=0.49\linewidth]{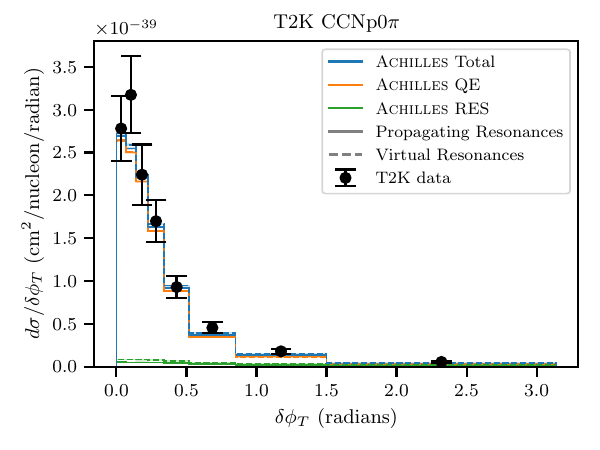}
    \caption{
    \textbf{Left:}
        Comparison of the CC$1\pi^{+}$ differential cross section with respect to the transverse boosting angle $\delta \alpha_T$ measured by T2K~\cite{T2K:2021naz} compared to predictions by \achilles.
    \textbf{Right:}
        Comparison of the CC$0\pi$ differential cross section with respect to the transverse deflecting angle $\delta\phi_T$ measured by T2K~\cite{T2K:2018rnz}. 
    }
    \label{fig:T2K_extra}
\end{figure}

This appendix collects additional comparisons between experimental measurements and the predictions of \achilles.
\Cref{fig:MINERvA_extras} shows a comparison to the CC$0\pi$ differential cross sections measured by MINER$\nu$A~\cite{MINERvA:2018hba} with respect to several different kinematic variables.
\Cref{fig:T2K_extra} shows a comparison of measurements by T2K.

\twocolumngrid

\bibliographystyle{apsrev4-1.bst}
\bibliography{bibliography}

\end{document}